\documentclass[prd,superscriptaddress,nofootinbib,preprint]{revtex4-1}
\usepackage[utf8]{inputenc}
\usepackage{graphicx}
\usepackage{amsmath}
\usepackage{tabularx}
\usepackage{mathtools}
\usepackage{cancel}
\usepackage{color}
\usepackage{subcaption}
\usepackage{caption}
\usepackage{slashed}
\usepackage{array}
\usepackage{float}
\usepackage{ amssymb }
\usepackage[colorlinks=true,citecolor=darkred,urlcolor=darkred, pdfborder={0 0 0}]{hyperref}
\definecolor{linkcolor}{rgb}{0,0,0.5}
\definecolor{darkred}{rgb}{0.6,0,0}

\usepackage{geometry}
\usepackage{siunitx} % Required for alignment
\usepackage{multirow} % Required for multirows
\DeclarePairedDelimiter\abs{\lvert}{\rvert}%
\usepackage{setspace}

\def\non{\nonumber}

\begin{document} 

\title{Probing Doubly and Singly Charged Higgs  at  $pp$ Collider  HE-LHC}
\author{Rojalin Padhan }\email{rojalin.p@iopb.res.in}
\affiliation{Institute of Physics, Sachivalaya Marg, Bhubaneswar 751005, India}
\affiliation{Homi Bhabha National Institute, BARC Training School Complex,
Anushakti Nagar, Mumbai 400094, India}
\author{Debottam Das }\email{debottam@iopb.res.in}
\affiliation{Institute of Physics, Sachivalaya Marg, Bhubaneswar 751005, India}
\affiliation{Homi Bhabha National Institute, BARC Training School Complex,
Anushakti Nagar, Mumbai 400094, India}
\author{Manimala Mitra}\email{manimala@iopb.res.in}
\affiliation{Institute of Physics, Sachivalaya Marg, Bhubaneswar 751005, India}
\affiliation{Homi Bhabha National Institute, BARC Training School Complex,
Anushakti Nagar, Mumbai 400094, India}
\author{Aruna Kumar Nayak}\email{nayak@iopb.res.in}
\affiliation{Institute of Physics, Sachivalaya Marg, Bhubaneswar 751005, India}
\affiliation{Homi Bhabha National Institute, BARC Training School Complex,
Anushakti Nagar, Mumbai 400094, India}
\preprint{\textbf{IP/BBSR/2019-5}}
 %%%%%%%%%%%%%%%%%%%%%%%%%%%%%%%%%%%%%%%%
\bibliographystyle{unsrt} 
%%%%%%%%%%%%%%%%%%%%%%%%%%%%%%%%%%%%%%%%%%%%%
\begin{abstract}
We analyse the  signal sensitivity of multi-lepton final states at  collider that can arise from doubly and singly charged Higgs decay in a type-II seesaw framework. We assume triplet vev to be very small and degenerate  masses for both the charged Higgs states. The leptonic branching ratio of doubly and singly charged Higgs states have a large dependency on the neutrino oscillation parameters, lightest neutrino mass scale, as well as neutrino mass hierarchy. We explore this as well as the relation between the leptonic branching ratios of the singly and doubly charged Higgs states in detail. We evaluate the effect of these uncertainties on the production cross-section of multi-lepton signal. Finally, we present a detailed analysis of multi-lepton final states for a future hadron collider HE-LHC, that can operate with  center of mass energy $\sqrt{s}=27$ TeV.

\end{abstract}
\vspace{-3cm}
\maketitle
\section{Introduction}
The discovery of the Higgs boson at the Large Hadron Collider (LHC) has experimentally proven that  fermions and gauge bosons masses in the Standard Model (SM) are generated via  Brout-Englert-Higgs (BEH) mechanism. However, one of the key questions that still remains unexplained   is the origin of light neutrino masses and mixings.  A number of neutrino oscillation experiments have  observed that, the solar and atmospheric neutrino mass splittings are $\Delta m^2_{12} \sim 10^{-5}$ $ \rm{eV}^2$ and $\Delta m^2_{13} \sim 10^{-3}$ $ \rm{eV}^2$, and the mixing angles are $\theta_{12} \sim 32^\circ$, $\theta_{23} \sim 45^\circ$, and 
$\theta_{13} \sim 9^\circ$ \cite{deSalas:2017kay}. A Dirac mass term of the SM neutrinos can be generated by extending the SM to include right-handed neutrinos.  However, this  requires very small Yukawa couplings, that introduces $\mathcal{O}(10^{-11})$ order of magnitude hierarchy between SM fermion Yukawa couplings, and hence is unappealing. A different ansatz is that neutrinos are their own anti-particles and hence, their masses can have a different origin compared to the other  SM fermions. One of such  profound mechanisms is seesaw, where tiny eV masses of the Majorana neutrinos  are generated from  lepton number violating (LNV) $d=5$ operator $LLHH/\Lambda$~\cite{Weinberg:1979sa,Wilczek:1979hc}. Being, a higher dimensional 
non-renormalizable operator,  there can be different UV completed theories behind this operator, commonly known as, type-I, -II, and -III seesaw mechanisms. These models include extensions of the SM fermion/scalar contents by SM singlet fermions~\cite{Minkowski:1977sc,Mohapatra:1979ia,Yanagida:1979as,GellMann:1980vs,Schechter:1980gr,Babu:1993qv,Antusch:2001vn}, $SU(2)_L$ triplet scalar boson~\cite{Magg:1980ut,Cheng:1980qt,Lazarides:1980nt,Mohapatra:1980yp}, and $SU(2)_L$ triplet fermion~\cite{Foot:1988aq}, respectively.\\

Among the above,  type-II seesaw model, where a triplet scalar field with the hypercharge $Y=+2$ is added to the SM, has  an extended scalar sector. There are seven physical Higgs states that includes singly and doubly charged Higgs, CP even and odd neutral Higgs. The details of the Higgs spectra have been discussed in \cite{Arhrib:2011uy,Dev:2013ff}. The neutral component of the triplet acquires a vacuum expectation value (vev) $v_{\Delta}$, and generates neutrino masses through the Yukawa interactions. The same Yukawa interaction between the lepton doublet and the triplet scalar field  also dictates the  charged Higgs phenomenology in this model. The presence of a doubly charged Higgs ($H^{\pm\pm}$) is the most appealing feature of this model, and hence,  a  discovery of this exotic particle will be a smoking gun signature of type-II seesaw. \\

A number of searches  have already been performed to search for the signatures of the doubly charged Higgs (see  \cite{Akeroyd:2005gt} for Tevatron,  and  \cite{Perez:2008ha,Melfo:2011nx,delAguila:2008cj,Chakrabarti:1998qy,Aoki:2011pz,Akeroyd:2011zza,Chun:2013vma,delAguila:2013mia, Banerjee:2013hxa,kang:2014jia,Han:2015hba,Han:2015sca,Babu:2016rcr,Du:2018eaw,Antusch:2018svb} for LHC).  Depending on the triplet vev, the doubly-charged Higgs boson can have distinct   decay modes. For low vev $v_{\Delta} \lesssim 10^{-4}$ GeV,  this can decay into same-sign di-lepton, whereas,  for $v_{\Delta} \geq 10^{-4}$ GeV,  this can decay to  same-sign gauge bosons. For non-degenerate masses of  doubly and singly charged Higgs, another  possible decay is the   cascade decay of a doubly charged Higgs to a singly charged Higgs and SM states. This has been explored in \cite{Perez:2008ha,Melfo:2011nx,delAguila:2008cj}.  The CMS and ATLAS collaboration have searched for the same-sign di-lepton final states 
with different flavors, and excluded the mass of the doubly-charged Higgs ($M_{H^{\pm \pm}})$ below 820 and 870 GeV, respectively, at  95$\%$ C.L. \cite{Aaboud:2017qph, CMS-PAS-HIG-16-036}. An alternative search where the $H^{\pm \pm}$ is produced in association with two jets, i.e.,   vector boson fusion  gives relaxed constraints \cite{Khachatryan:2014sta, Sirunyan:2017ret}. Another scenario where doubly-charged Higgs decays to same-sign $W^\pm$ boson pairs. The collider signatures and the discovery prospects of this scenario have been discussed in \cite{Kanemura:2013vxa,Kanemura:2014goa,Kanemura:2014ipa}, and \cite{Mitra:2016wpr, Ghosh:2017pxl}. ATLAS collaboration have searched for the same final state and excluded the doubly-charged Higgs mass between 200 and 220 GeV  at  95$\%$ C.L.~\cite{Aaboud:2018qcu}. Previous searches for $H^{\pm \pm}$ in the pair-production channel and their subsequent decays into same-sign leptons at LEP-II has put a constraint $M_{H^{\pm \pm}} > 97.3 $ GeV at $95 \%$ C.L. \cite{Abdallah:2002qj}. For  discussions on Higgs triplet model at a linear collider, see \cite{Shen:2015bna,Blunier:2016peh,Cao:2016hvg,Guo:2016hjt,Agrawal:2018pci} and  at $ep$ collider, see~\cite{Dev:2019hev}. Displaced vertex signatures have been discussed in Ref.~\cite{Dev:2018kpa,Antusch:2018svb}. A review on this model is presented in ~\cite{Cai:2017mow}. 

 While a number of searches at the LHC are ongoing to experimentally verify the presence of the doubly-charged Higgs boson, in this work we explore the impact of light neutrino mass hierarchy, neutrino oscillation parameters, as well as, the lightest neutrino mass scale $m_0$ on  $H^{\pm \pm} $  searches. We  relate the  branching ratios of doubly and singly charged Higgs decays for both normal and inverted mass hierarchy. We find that among the different leptonic modes, the decay mode of doubly charged Higgs into two same-sign electron, and the decay mode of a singly charged Higgs into an electron and neutrino are the least uncertain for inverted neutrino mass  ordering, and has the potential to differentiate neutrino mass hierarchy. We also discuss how the inclusion of uncertainties in the neutrino oscillation parameters affect the  theory cross-section, which may in turn change the mass limits of doubly charged Higgs in individual channel. As it is well known that for c.m. energy  $\sqrt{s}=13\ ( \text{or}\ 14)$ TeV LHC, production of multi-TeV  $H^{\pm\pm}$  will be difficult due to suppressed cross-section. However, increasing c.m. energy one can probe heavier $H^{\pm\pm}$. Therefore we consider pair-production and associated production of the doubly-charged Higgs boson and its subsequent decays into leptonic states, including tau's, and analyse the discovery prospects of doubly charged Higgs at a future hadron collider (HE-LHC), that can operate  with  c.m. energy  $\sqrt{s}=27$ TeV. We consider both the tri and four lepton final states, and present a detail analysis taking into account different possible SM background processes. We find that in addition to the associated production, the pair-production of doubly charged Higgs also gives  a significant contribution to the tri-lepton final states. We consider a wide range of doubly charged Higgs mass, and explore the sensitivity reach with   the projected luminosity (15 $\rm{ab}^{-1}$) of HE-LHC~\cite{Abada2019,Cepeda:2019klc}.
 
Our paper is organized as follows: we briefly review the basics of the type-II seesaw model in Sec.~\ref{model}. In Sec.~\ref{br}, we discuss leptonic branching ratios of doubly-charged ($H^{\pm\pm}$) and singly charged ($H^{\pm}$) Higgs, and the  relation between $H^{\pm\pm}$ and $H^{\pm}$ decays. In Sec.~\ref{limit}, we discuss the effect of uncertainties in neutrino oscillation parameters on the  production cross-section of multi-lepton signal. In Sec.~\ref{multilep},  we present the simulation of multilepton signal at $\sqrt{s}=27$ TeV LHC. Finally, we present our conclusions in Sec.~\ref{conclu}.
\section{Model description \label{model}}	

In this section,  we briefly discuss    type-II seesaw model \cite{Magg:1980ut, Cheng:1980qt, Lazarides:1980nt, Mohapatra:1980yp}.  The  model is based on the gauge group as of the SM gauge group, \ $G_{SM} =\ SU(3)_C \times \ SU(2)_L \times \ U(1)_Y$.  Apart from the SM particles,  the particle spectrum also contains {one additional }  $SU(2)_L $ triplet scalar  $\Delta$ with hypercharge $\ \rm{Y_\Delta} = +2$ : 	
\begin{eqnarray}
\Delta=\begin{pmatrix} \frac{\Delta^+}{\sqrt{2}} & \Delta^{++} \\ \frac{1}{\sqrt{2}}(v_{\Delta}+\delta^0+i\eta^0) & -\frac{\Delta^+}{\sqrt{2}}
\end{pmatrix}.
\end{eqnarray}
{The SM Higgs doublet is represented as follows, }
\begin{eqnarray}
\Phi= \begin{pmatrix} \phi^+ \\ \frac{1}{\sqrt{2}}(v_{\phi}+\phi^0+i\chi^0)  \end{pmatrix} . 
\end{eqnarray}
After electroweak symmetry breaking, the real part of neutral Higgs $\phi^0$ and $\delta^0$ acquire  vevs,  denoted as $v_{\phi}$ and $v_{\Delta}$,  respectively. The two vevs satisfy $v^2=v^2_{\phi}+v^2_{\Delta}=(246 \, \, \rm{GeV})^2$.  {Below, we discuss different terms of the Lagrangian. }
\begin{itemize}
\item The kinetic Lagrangian for the scalar sector is,
\begin{eqnarray} 
{\cal L}_{\text{kin}} & = & \rm{(D_\mu \Phi)^\dagger (D^\mu \Phi)} \ + \  \rm{Tr} [(D_\mu \Delta)^\dagger (D^\mu \Delta)].
\end{eqnarray}
The covariant derivatives in Eq. 3 are defined as, 
\begin{eqnarray}
\rm{D_\mu\Phi}=\rm{\partial_\mu\Phi+i\frac{g}{2}\tau^aW_\mu^a\Phi}+\rm{i g' \frac{Y_\Phi}{2}B_\mu\Phi},
\end{eqnarray}
\begin{eqnarray}\rm{D_\mu \Delta}=\rm{\partial_\mu \Delta+i\frac{g}{2}[\tau^aW_\mu^a,\Delta]}+\rm{i g' \frac{Y_\Delta}{2}B_\mu\Delta}.
\end{eqnarray}
Both $v_\phi \ \text{and} \ v_\Delta$ contribute to the masses of weak gauge bosons at tree level. {Therefore,} the $\rho$ - parameter (=$\frac{M_W^2}{M_Z^2 \cos^2\theta_W}$) in this model is given by, 
\begin{eqnarray}
\rho = \frac{1 + \frac{2 v_\Delta^2}{v_\phi^2}}{1 + \frac{4 v_\Delta^2}{v_\phi^2}}.
\end{eqnarray} The current electroweak precision data~\cite{Patrignani:2016xqp} gives the value of $\rho$ parameter,  $\rho=1.00037\pm0.00023$, which is $ 1.6  \sigma $ away from the  tree-level SM prediction. We consider $ 2.18 \sigma $ experimental error on the measured central value of $\rho$ parameter and estimate a conservative bound on $v_\Delta, \ \text{i.e.},  \  v_\Delta \lesssim 2$  GeV. Thus the two vevs 
satisfy $v_\Delta \ll v_\phi$.
\item The Yukawa Lagrangian of this model  is given by,	
\begin{eqnarray}\rm{ {\cal L}_{Y}(\Phi, \Delta)}&= {{\cal L}_{Y}^{SM}(\Phi)}& \ + { Y^\nu \ L_L^T \ C \ i \sigma_2 \ \Delta \ L_L \ }+\rm{h.c.}
\end{eqnarray}

Here, the first term in ${{\cal L}_{Y}(\Phi,\Delta)}$ represents the Yukawa interactions of  the SM Higgs doublet ($\Phi$) and the second term is the needed Yukawa interaction of  the triplet Higgs ($\Delta$), that generates neutrino mass. ${Y^\nu}$ is Yukawa coupling  matrix, C is the charge conjugation operator, and $\sigma_2$ is the Pauli matrix. ${L_L}$ is the left chiral lepton doublet. Once,  the triplet Higgs ($\Delta$) acquires   vacuum expectation value $v_\Delta$, the second term in 	${ {\cal L}_{Y}(\Phi, \Delta)}$ generates a Majorana mass for neutrino, which is given by, 
\begin{eqnarray}
M^{\nu}= \sqrt{2} \ Y^\nu \ v_{\scriptscriptstyle\Delta}.
\end{eqnarray} 
In the above  $M^{\nu}$  is a complex symmetric $3 \times 3 $ matrix, which can be diagonalized by  an unitary transformation defined as $M^\nu = V^*_{\rm{PMNS}} m^\nu_{d} V_{\rm{PMNS}}^\dagger$ . Here $m^\nu_d = \text{diag} (m_1,\ m_2,\ m_3)$,   is diagonal light neutrino mass matrix,  and $V_{\rm{PMNS}}$ is the neutrino mixing matrix parametrised  by the  three mixing angles ($\theta_{12}, \ \theta_{13},\ \theta_{23}$) and three phases ($\phi_1, \ \phi_2,\ \delta $).

\item The scalar potential \cite{Arhrib:2011uy} with the two Higgs fields $\Phi$ and $\Delta$ is  
\begin{eqnarray}
V(\Phi,\Delta)&=&m^2\Phi^\dagger\Phi +   M^2\rm{Tr}(\Delta^\dagger\Delta)+\left(\mu \Phi^Ti\sigma_2\Delta^\dagger \Phi+\rm{h.c.}\right)+\frac{\lambda}{4}(\Phi^\dagger\Phi)^2 \nonumber\\
&+&\lambda_1(\Phi^\dagger\Phi)\rm{Tr}(\Delta^\dagger\Delta)+\lambda_2\left[\rm{Tr}(\Delta^\dagger\Delta)\right]^2+\lambda_3\rm{Tr}[(\Delta^\dagger\Delta)^2]+\lambda_4\Phi^\dagger\Delta\Delta^\dagger\Phi. ~~~~~
\end{eqnarray}
All operators in the above scalar potential are self conjugate except the operator containing $\mu$. Therefore,  all parameters except $\mu$ are real. Although $\mu$ can pick up a would-be CP phase, this phase is unphysical and can always be absorbed in a redefinition of the scalar fields. Together $Y^{\nu}$ and the  $\mu$ term violate  lepton number symmetry  in this model. Minimization of $	\rm{V(\Phi,\Delta)}$ gives the following two  conditions \cite{Arhrib:2011uy}:
\begin{eqnarray} M^2 &= &\ \frac{2 \mu v_{\phi}^2 - \sqrt{2}(\lambda_1 + \lambda_4)v_{\phi}^2 v_\Delta - 2\sqrt{2}(\lambda_2 + \lambda_3) v_{\Delta}^3}{ 2\sqrt{2} v_\Delta},\\
m^2 &=& \frac{\lambda v_{\phi}^2}{4} -  \sqrt{2} \mu v_\Delta + \frac{(\lambda_1 + \lambda_4)v_{\Delta}^2}{2}. 
\end{eqnarray}
Thus the two mass parameters $m^2$ and $M^2$ can be eliminated which leaves 8 free parameters ($v_\Delta,v_\phi,\mu,\lambda,\lambda_{1},\lambda_{2},\lambda_{3},\lambda_{4}$). Further $v^2 \equiv v^2_{\Phi}+v^2_{\Delta}=(246 \, \, \rm{GeV})^2$, reduces this set of free parameters down to seven. 
{There are  ten real scalar degrees of freedom present in this model, out of which  three are the would be  Goldstone bosons, and they  give masses to the SM weak gauge bosons after electroweak symmetry breaking. The remaining seven states are the physical Higgs bosons. Doubly charged scalars, $\Delta^{\pm \pm} (\equiv H^{\pm\pm})$  is purely triplet, and is already in mass eigenbasis. The singly charged scalars ($\phi^{\pm}$,~$\Delta^{\pm}$)  and neutral scalars 
($\chi^{0},\eta^0,\phi^0,\delta^0$) are not physical fields, as they share non-trivial mixings among them.  We denote the mass eigenstates of the singly charged scalars by  $G^{\pm}$ and $H^{\pm}$, that are linear combinations of  $\phi^{\pm}$ and $\Delta^{\pm}$. Similarly, the two CP-odd physical  fields are denoted by  $G^{0}$ and $A$ (linear combinations of  $\chi^{0}$ and $\eta^{0}$)}. The SM Higgs field ($h$) and a heavy Higgs ($H$) are massbasis of the two neutral CP-even states $\phi^{0}$ and $\delta^{0}$. $G^\pm$ and $G^{0}$ are the three Goldstone bosons. These scalar mixings  are small,  as they are  proportional to the triplet vev $(v_\Delta)$. The presence doubly charged Higgs ($H^{\pm\pm}$) is the unique feature of this model. For detail discussion on mass of these scalars and doublet triplet mixing angles, see  \cite{Arhrib:2011uy}.

Assuming  $v_\Delta \ll v_\phi$, the masses of the physical Higgs bosons   are given by \cite{Arhrib:2011uy},
\begin{eqnarray}
M_{H^{\pm\pm}}^2 \simeq M_{\Delta}^2-\frac{\lambda_4}{2}v_\phi^2, \ M_{H^\pm}^2 \simeq M_\Delta^2-\frac{\lambda_4}{4}v_\phi^2, \ 
M_h^2 \simeq 2 v_\phi^2\lambda, \ 
M_H^2 = M_A^2 \simeq M_{\Delta}^2 \nonumber, 
\end{eqnarray}

where  $M_\Delta^2 \equiv \frac{\mu v_{\phi}^2}{\sqrt{2}v_\Delta}$. {We identify the $h$ field as the neutral SM Higgs, with its mass denoted as  $M_h$. The mass of the SM Higgs  is primarily governed by $\lambda$.}
The parameter $M_\Delta$  determines  the mass scale of all other Higgs bosons. %$\lambda$ can be determined from $M_h$  and vevs.
Mass square differences between the scalars are given by 
\begin{eqnarray}
M_{H^{\pm}}^2 - M_{H^{\pm\pm}}^2 \simeq \frac{\lambda_4}{2}v_\phi^2, \
M_{H/A}^2 - M_{H^{\pm}}^2 \simeq \frac{\lambda_4}{4}v_\phi^2 . 
\end{eqnarray}
Note that,  the quartic coupling $\lambda_4$ of  the potential  {dictates}   the mass splitting between $H^{\pm}-H^{\pm\pm}$ and $H(A)-H^{\pm}$. These two mass square differences are of similar order. 
Taking into account the electroweak precision data \cite{Lavoura:1993nq}, the mass 
 splitting of triplet Higgs is constrained as  $\delta M< $40$ \, \rm{GeV}$ \cite{Melfo:2011nx,Chun:2012jw}. Therefore,  the value of  $\lambda_4  $ defines  three different mass spectrum of the  triplet  Higgs,
\begin{itemize}
\item[$\bullet$] $\lambda_4 = 0 $ \ (Degenerate Scenario) : $M_{H^{\pm\pm}} \simeq M_{H^{\pm}} \simeq M_{H/A}$, 
\item [$\bullet$]$\lambda_4 > 0 $ \ (Positive  Scenario) : $M_{H^{\pm\pm}} < M_{H^{\pm}} < M_{H/A}$, 
\item[$\bullet$] $\lambda_4 < 0 $ \ (Negative  Scenario): $M_{H^{\pm\pm}} > M_{H^{\pm}} > M_{H/A}$. 
\end{itemize}
In our entire analysis,  we assume  degenerate scenario for triplet Higgs mass, where all the triplet like scalars have  same masses. The lightest neutral Higgs, that is primarily originated from the doublet $\Phi$ is considered as the SM like Higgs. In Degenerate Scenario, one triplet Higgs will not be able to decay into another triplet Higgs and a gauge boson. Going beyond Degenerate Scenario opens up a number of other decay possibilities,  such as the cascade decays $H^{++} \to H^+ W^{+\star} \ , \ H^+ \to H/A \ W^{+\star} $ in Negative Scenario and  $H^{+} \to H^{++} W^{-\star} \ , \ H/A \to H^+  W^{-\star} $ in Positive Scenario.  As discussed in \cite{Aoki:2011pz}, these decays can be dominant if the mass differences between the charged Higgs states, $\delta M > 1$ GeV. In other mass ranges,   these are  very suppressed. In the next section, we discuss the decay widths and branching ratios of different Higgs states, assuming a degenerate scenario. Therefore, cascade
decay is not very relevant in our analysis.

\end{itemize}
\begin{figure}
	
	\includegraphics[scale=1,width=7cm,height=6cm]{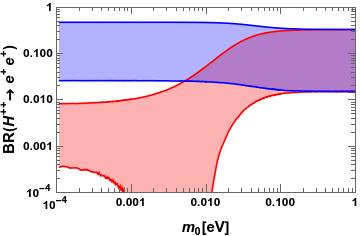}
	\includegraphics[scale=1,width=7cm,height=6cm]{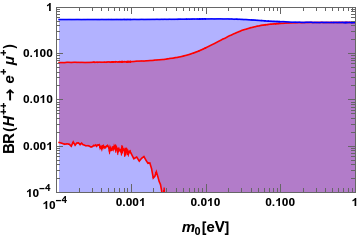}\\
	\includegraphics[scale=1,width=7cm,height=6cm]{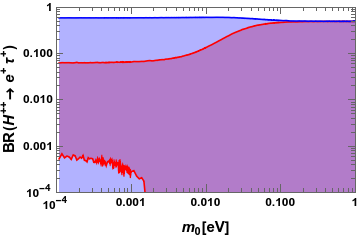}
	\includegraphics[scale=1,width=7cm,height=6cm]{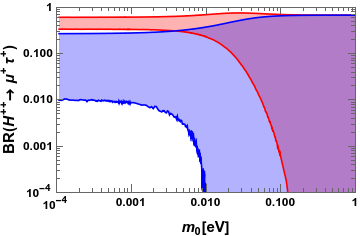}\\
	\includegraphics[scale=1,width=7cm,height=6cm]{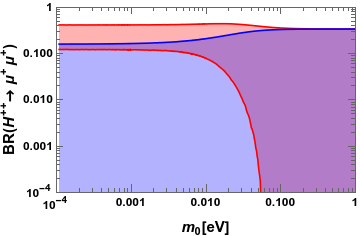}
	\includegraphics[scale=1,width=7cm,height=6cm]{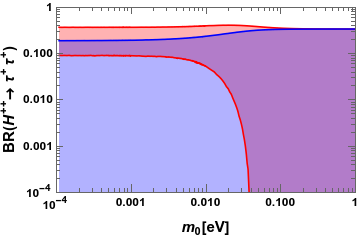}
	\caption{Variation of branching ratios of $ H^{++} \to l^{+} l^{+} $ (where $l=e, \mu,\tau$) as a function of lightest neutrino mass $m_0$ ($m_0$ is $m_1$ in NH and $m_3$ in IH). The band  represents the uncertainty in branching ratio due to  $3\sigma$ variation of  neutrino oscillation parameters. We vary the CP phases  (Dirac and Majorana phases)  in  between  $0-2\pi$. Blue (red) band represents IH (NH) of neutrino mass pattern.}
	\label{fig:hppbr}
\end{figure}
\section{Branching ratios of $H^{\pm \pm }$ and $H^{\pm}$ \label{br}}

The decay properties of charged Higgs states in different $v_{\Delta}$  region has been discussed extensively  in the literature~\cite{Perez:2008ha,Garayoa:2007fw}. 
 For $v_{\Delta}<10^{-4}$ GeV the dominant decay channel of the  doubly-charged Higgs is  $H^{\pm \pm}\to l_i^{\pm}l_j^{\pm}$  and singly-charged Higgs is $H^{\pm}\to l_i^{\pm}\nu$, which is clear from Figs. 4, 5 of ~\cite{Perez:2008ha}. In this region of $v_\Delta$ it will be possible to find out the correct neutrino mass ordering by measuring the leptonic branching ratios of the charged Higgs states \cite{Chun:2003ej}. 
 Note that, in the leptonic channel, the same Yukawa coupling governs both the doubly-charged and singly-charged Higgs decays. Therefore,  the leptonic decays of these two Higgs states are related. Below, we discuss the different decay channels and the relation between $H^{\pm\pm}$ and $H^{\pm}$ decays in detail. 
\begin{itemize}
\item  $\boldsymbol{H^{\pm \pm}}$ \textbf{Decays} \\

 Partial decay width of $H^{\pm\pm}$ to a pair of same-sign leptons  ~\cite{Perez:2008ha} is given by 
\begin{equation}
\Gamma_{l_il_j} \equiv \Gamma(H^{\pm \pm}\to l_i^\pm l_j^\pm) = \frac{1}{4 \pi(1+\delta_{ij})}\abs{Y^\nu_{ij}}^2 M_{H^{\pm \pm}}.
\end{equation} 
 We consider $v_{\Delta}<10^{-4}$ GeV, and hence,   $H^{\pm \pm}$ predominantly  decays  to leptonic final states. The 
decay branching ratio (BR) has the following form,
\begin{equation}
{\rm{BR}}(H^{\pm \pm}\to l_i^\pm l_j^\pm) =  \frac{\Gamma_{l_il_j}}{\sum_{kl}	\Gamma_{l_kl_l}} = \frac{2}{(1+\delta_{ij})} \frac{\abs{Y^\nu_{ij}}^2}{\sum_{kl}\abs{Y^\nu_{kl}}^2 }, \end{equation} 
where 
\begin{equation}
\quad \sum_{kl}\abs{Y^\nu_{kl}}^2 = \dfrac{1}{2v^2_\Delta}\sum_i m_i^2.
\end{equation}

 In Fig.~\ref{fig:hppbr}, we plot the ${\rm{BR}}(H^{\pm \pm}\to l_i^\pm l_j^\pm)$ as a function of lightest neutrino mass  ($m_0$) for both normal (NH) and inverted (IH) mass hierarchy. Blue and red bands represent IH and NH,  respectively.  Similar plots have already been presented in ~\cite{Perez:2008ha,Garayoa:2007fw} considering  the  $3\sigma$ range of neutrino mixing angles and mass square differences. But here we  also vary all phases in between $0-2\pi$ and we consider current value of neutrino oscillation parameters~\cite{deSalas:2017kay}, including nonzero $\theta_{13}$. Some notable points about these plots are as follows:
 \begin{enumerate}
	\item  $m_{0}>0.1$ eV represents the quasi-degenerate neutrino mass spectrum, which is disallowed by cosmological data \cite{Aghanim:2018eyx}.
   \item  For $m_{0}<0.1$ eV,  and for the modes  $ e^\pm e^\pm, e^\pm \mu^\pm, e^\pm \tau^\pm$, the maximum  value of the branching ratio in IH is larger than that in NH. For  $\mu^{\pm}\mu^{\pm},\mu^{\pm}\tau^{\pm},\tau^{\pm}\tau^{\pm}$ mode,  it is the reverse. This behaviour can be understood from Eq.~\ref{eq_emu},~\ref{eq_ee},~\ref{eq_mumu}, which are the ratios between maximum branching ratio in IH and NH for a given decay channel in the hierarchical regime with $m_0\approx0$. The exact equations are presented in Sec.~\ref{app}.
   \begin{equation}
   \dfrac{\text{BR}^{\max}(H^{\pm \pm }\to e^{\pm } \mu^{\pm })_{\text{IH}}}{\text{BR}^{\max}(H^{\pm \pm }\to e^{\pm } \mu^{\pm })_{\text{NH}}}
   \approx\dfrac{c_{13}^2 \left(c_{12}^2 s_{13} s_{23}+2 c_{23} c_{12} s_{12}-s_{12}^2 s_{13} s_{23}\right)^2}{2 \left(0.2 c_{12} c_{23} s_{12}+s_{13} s_{23} \left(0.2  s_{12}^2+1\right)\right){}^2}\approx6.6, \label{eq_emu}
   \end{equation}
    \begin{equation}\dfrac{\text{BR}^{\max}(H^{\pm \pm }\to e^{\pm } e^{\pm })_{\text{IH}}}{\text{BR}^{\max}(H^{\pm \pm }\to e^{\pm } e^{\pm })_{\text{NH}}}
   \approx\dfrac{ c_{13}^4}{2\left(0.2 c_{13}^2 s_{12}^2+s_{13}^2\right)^2}\approx50 ,\label{eq_ee}
   \end{equation}
   \begin{equation}
   \dfrac{\text{BR}^{\max}(H^{\pm \pm }\to \mu^{\pm } \mu^{\pm })_{\text{IH}}}{\text{BR}^{\max}(H^{\pm \pm }\to \mu^{\pm } \mu^{\pm })_{\text{NH}}}\approx \dfrac{c_{23}^4}{2 (0.2 c_{12}^2 c_{23}^2 +c_{13}^2 s_{23}^2)^2}\approx0.45 . \label{eq_mumu}
   \end{equation}
  
   In the above, we consider the values of oscillation parameters, that  maximize the numerator and denominator separately, as we are interested in the relative  comparison of maximum branching ratios in NH and IH. The approximate 
   expressions in the above equations clearly show that for IH neutrino mass spectrum, $e^{\pm}e^{\pm}$ and $e^{\pm}\mu^{\pm}$ final states will be more favourable, as these channels can have large branching ratios. Although the final state $e^{\pm}\tau^{\pm}$ has large branching, however, further leptonic decays of $\tau^{\pm}$ will give suppression in cross-section.
   
   	\item There exist a large uncertainty in branching ratios, that somewhat reduces for the choice of  CP phases to be zero. Among the different leptonic modes,  $H^{\pm \pm }\to e^\pm e^\pm $ in IH is the most favourable mode  for the entire range of $m_0$,  as this decay mode has a less uncertainty in branching ratio, and there is a definite predicted lower value of ${\rm{BR}}(H^{\pm \pm} \to e^{\pm} e^{\pm})$. Irrespective of the  value of lightest neutrino mass, and the variation of oscillation parameters, the discovery of $H^{\pm \pm}$ will therefore  be  more favourable  in this channel. An observation of $H^{\pm\pm}$ in any other leptonic decay mode except $e^{\pm}e^{\pm}$ mode with a branching ratio limit ${{\rm{BR}}}(H^{\pm \pm} \to e^{\pm} e^{\pm})< 0.015$  will indicate  normal mass hierarchy in the light neutrino sector. 
   	
    \item Note that, except $H^{\pm \pm} \to e^{\pm} e^{\pm}$ in IH, all other decay modes heavily depend on the oscillation parameters, and $m_0$. Moreover, for those decays, there may exist a cancellation region, in which the branching ratio becomes highly suppressed. This occurs when different terms in the partial decay widths cancel out each other. This is to note that, for $H^{\pm \pm} \to e^{\pm} e^{\pm}$ in IH, such cancellation regions do not exist. As an example, the cancellation region for $H^{\pm \pm} \to e^{\pm} e^{\pm}$ in NH, that exists in between  $10^{-3} \  \text{eV}\lesssim m_0\lesssim10^{-2} \ \text{eV}$ can be explained as follows:     
   \text{    } For the choice  $m_1=10^{-3}$ eV, the largest neutrino mass  $ \sum_{i}m_i^2\approx m_3^2 \approx 4 \times 10^{-3}\ \text{eV}^2$. Considering the CP phases,   $\phi_1=2\delta-\phi_2=\pi$, one obtains, 
   	\begin{equation}
   	{{\rm{BR}}}(H^{\pm \pm} \to e^{\pm} e^{\pm})_{\text{NH}}\approx  10^{-6} \dfrac{(-\  c_{12}^2+8  s_{12}^2 c_{13}^2- 60 s_{13}^2)^2}{4 \times 10^{-3}}\approx 10^{-4}.
   	\end{equation}
   	The branching ratio in IH, is instead significantly large for the above choice of parameters.  For similar values of  $m_0=m_3=10^{-3} \ \text{eV},\phi_1,\phi_2  \ \text{and} \ \delta$ as mention in case of NH, one obtains,  \begin{equation}
   	{{\rm{BR}}}(H^{\pm \pm} \to e^{\pm} e^{\pm})_{\text{IH}}\approx 10^{-6}  \dfrac{(-60  \  c_{12}^2+ 60 \ s_{12}^2 c_{13}^2- s_{13}^2)^2}{ 7\times 10^{-3}}\approx 10^{-2}.
   	\end{equation}      
   	\item  For  NH scenario,  $H^{\pm \pm}\to \mu^\pm \mu^\pm /\mu^\pm \tau^\pm /\tau^\pm \tau^\pm$ channels have least uncertainty  for   $m_{0}<0.01$ eV, and hence the discovery of $H^{\pm \pm}$ into these above mentioned final states are more favourable for NH with $m_0< 0.01$ eV. Due to further decay of  $\tau$ into leptonic states, that involves smaller branching ratio, the overall cross-section in the channel with $\tau$ will be relatively smaller than the channel with $\mu \mu$. Furthermore, a doubly charged Higgs can not be fully reconstructed with the channel involving  leptons from $\tau$, due to the presence of missing energy. Therefore,  $H^{\pm \pm}\to \mu^\pm \mu^\pm$ decay mode will be more effective compare to other two $H^{\pm \pm}\to \mu^\pm \tau^\pm /\tau^\pm \tau^\pm$.
 \end{enumerate}
 
  \text{   } As we will discuss in the next section, the variation of decay branching ratios of $H^{\pm \pm} $  with oscillation parameters, as well as, the dependency on neutrino mass hierarchy  have large effect on the theory cross-section of the four-lepton final states.
\pagebreak
 \item  $\boldsymbol{H^{\pm}}$ \textbf{Decay} \\
 
  $H^{\pm}$ decays predominantly to a lepton and neutrino for $v_{\Delta} < 10^{-4}$ GeV. The partial decay width of $H^{\pm}$ to a lepton and neutrino ~\cite{Perez:2008ha} is given by, 
 \begin{equation}
 \Gamma_{l_j\nu_i} \equiv \Gamma(H^{\pm}\to l_j^\pm \nu_i) = \frac{1}{16 \pi}\abs{Y_{ij}^+}^2 M_{H^{\pm}}.
 \end{equation}  
 In the above,  $Y^+ = \cos\theta^+ \frac{m^\nu_{d}V_{PMNS}^\dagger}{v_\Delta}$ , $\theta^+$ is the singly charged Higgs mixing angle. For $v_{\Delta}<10^{-4}$ GeV,  branching ratio for the decay, $H^{\pm}\to l_j^\pm {\nu}_i$ is given by 
 \begin{equation}
 {\rm{BR}}(H^{\pm}\to l_j^{\pm}\nu_i) =  \frac{\Gamma_{l_j\nu_i}}{\sum_{kl}	\Gamma_{l_k\nu_{l}}} = \frac{\abs{Y_{ij}^+}^2}{\sum_{kl}\abs{Y_{kl}^+}^2 },
 \end{equation}
  where 
  \begin{equation}
  \sum_{kl}\abs{Y_{kl}^+}^2 = \dfrac{\cos^2\theta^+}{v^2_\Delta}\sum_i m_i^2.
 \end{equation} 
 
 \begin{figure}[b]
 	%\captionsetup{justification=centering,,margin=0.1cm}
 	\includegraphics[scale=1,width=7cm,height=6cm]{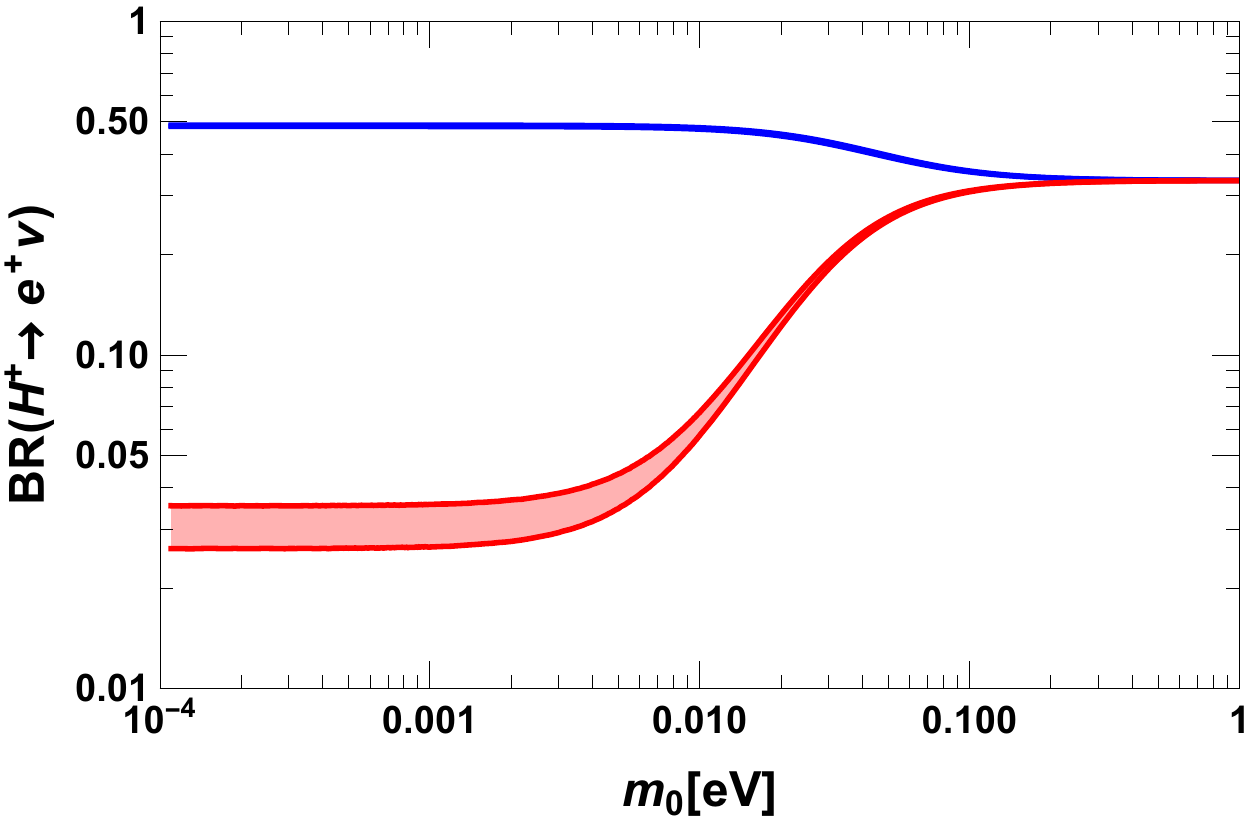}
 	\includegraphics[scale=1,width=7cm,height=6cm]{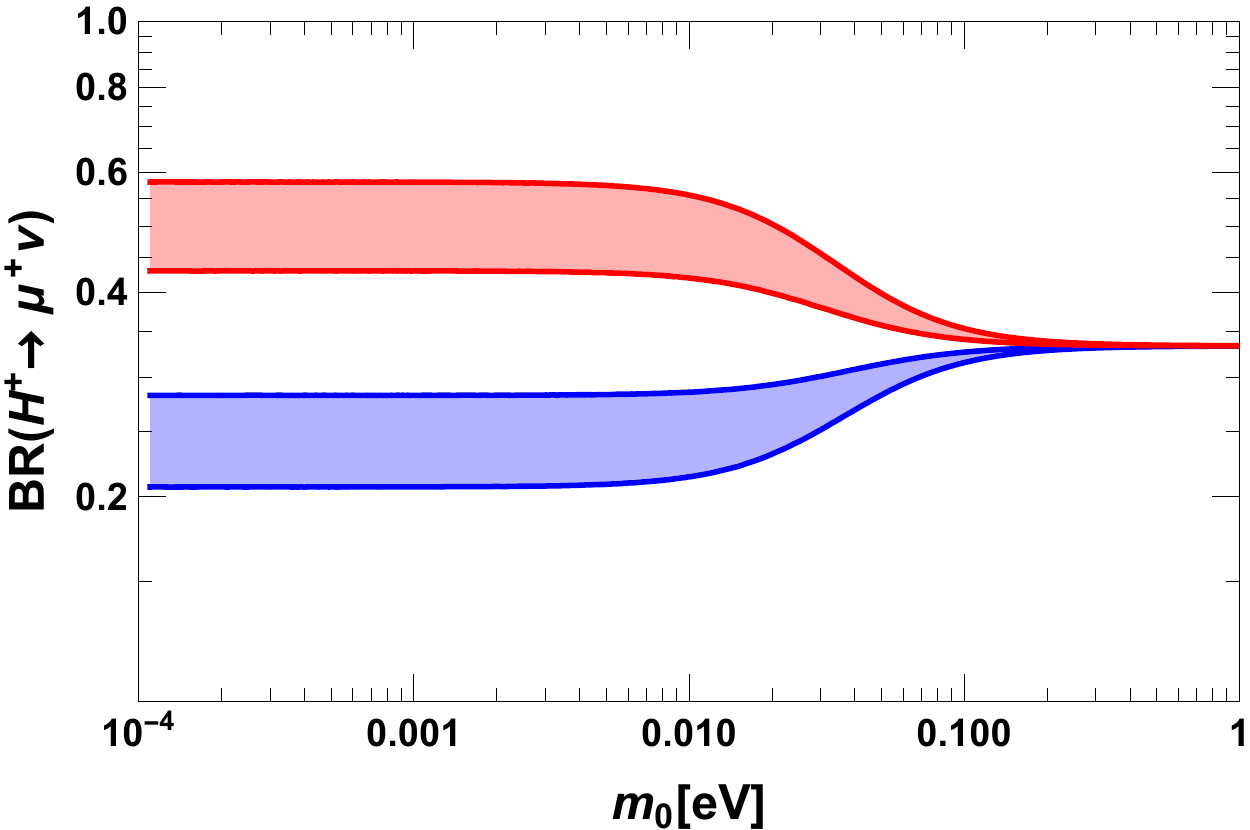}
 	\includegraphics[scale=1,width=7cm,height=6cm]{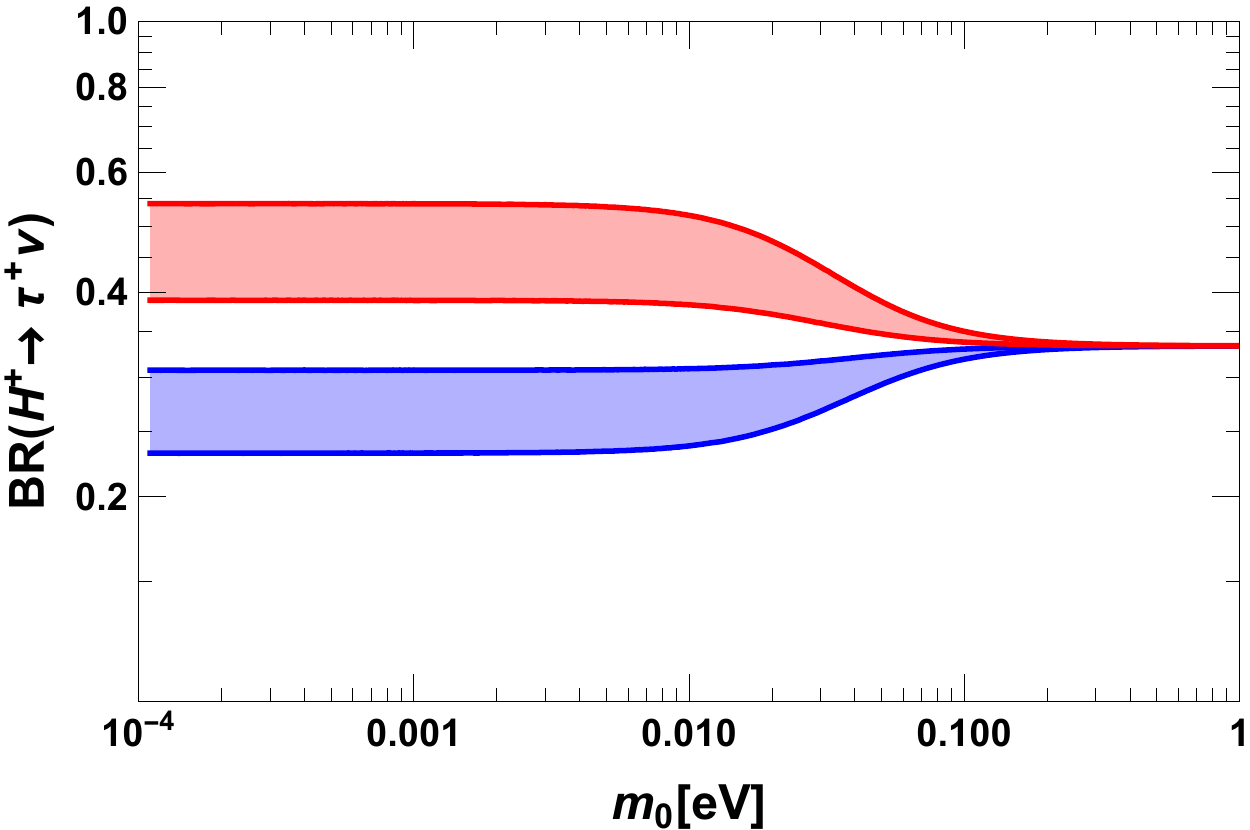}
 	\caption{Branching ratios of $ H^{+} \to l^{+} \nu $ (where $l=e, \mu,\tau$). Blue(red) band represents IH(NH) of neutrino mass pattern .}
 	\label{fig:hpbr}
 \end{figure}

  In Fig.~\ref{fig:hpbr}, we plot ${\rm{BR}}(H^{\pm}\to l_j^\pm \nu) \equiv \sum_i {\rm{BR}}(H^{\pm}\to l_j^\pm \nu_i)$ as a function of lightest neutrino mass ${m_0}$, where we consider  $3\sigma$ variation of neutrino oscillation parameters, and variation of  CP phases between  $0-2\pi$. Important points to be noticed are:
  \begin{enumerate}
 \item Note that, for   $m_{0} < 0.1$ eV and for IH, $ H^{\pm} \to e^{\pm} \nu $ has a large branching ratio ($\sim$ 0.5).  This decay channel is however,  has a smaller branching ratio for NH. In Sec.~\ref{app} (Appendix) this branching ratio has been calculated for $m_{0} \approx 0$. Maximum possible value of $\text{BR}(H^{\pm}\to e^{\pm } \nu)$ in IH compare to that in NH is given by,
 \begin{equation}\dfrac{\text{BR}^{\max}(H^{\pm}\to e^{\pm } \nu)_{\text{IH}}}{\text{BR}^{\max}(H^{\pm}\to e^{\pm } \nu)_{\text{NH}}}
 \approx\dfrac {c_{13}^2} {2(0.04 c_{13}^2 s_{12}^2+s_{13}^2)}\approx13. \end{equation}
 
\item Another important point to be noticed, is that, for $H^{\pm}$  the uncertainty in branching ratio is less compare to that for $H^{\pm\pm}$. This occurs because the Yukawa couplings in case of $H^{\pm}$ decay are independent of two Majorana phases $\phi_1$ and $\phi_2$. This is evident from the equations given in Sec.~\ref{app}.
\item Among the three decay modes of $H^{\pm}$,  $ H^{\pm} \to e^{\pm} \nu $  has less uncertainty in the  branching ratio, as the respective Yukawa is independent of Dirac CP phase $\delta$ (see Sec.~\ref{app}). This branching ratio depends on $m_0, \ \theta_{12},\ \theta_{13}$. The other two branching ratios for the muon and tau decay modes  depend on $\theta_{23}$ and $\delta$ as well.
\item The uncertainty in  branching ratios  for  $ H^{\pm} \to \mu^{\pm} \nu $ and $ H^{\pm} \to \tau^{\pm} \nu $ are nearly equal. This is clear from the  top right and bottom plots of Fig.~\ref{fig:hpbr}, where both the blue bands (in case of IH) have similar spread. This feature also exists in case of NH. 
  \end{enumerate}
\hspace{0.5cm}Assuming 100$\%$ branching ratios in leptonic decays,  CMS and ATLAS searches have constrained $H^{\pm \pm}$ below 820 GeV and 870 GeV, respectively. This is evident from the above discussion,  that the branching ratio in any of the leptonic channels can not reach upto $100\%$. In the next section, we re-evaluate the production cross-section of four-lepton final state, originating  from pair-production of doubly charged Higgs,  for different leptonic channels, taking into account the uncertainties of branching ratios. As an example, we consider the decay channels $ H^{\pm \pm} \to e^{\pm} e^{\pm} / e^{\pm} \mu^{\pm} / e^{\pm} \tau^{\pm} $ in IH, as they offer largest  values of branching ratios compared to NH. Note that, the maximum value of branching ratio for the other three decay modes $\mu^{\pm} \tau^{\pm} , \mu^{\pm} \mu^{\pm}$ and $ \tau^{\pm} \tau^{\pm}$ are relatively  smaller in IH. 
We provide a sample benchmark point in Table.~\ref{fig:BP1}, that shows $e^{\pm}\mu^{\pm}$ and $e^{\pm}\tau^{\pm}$ has large branching ratios in IH as compared to the other modes. This is to clarify that simultaneously the decay modes can not have maximum branching ratios. For the estimation in   NH, we assume the decay modes  $ H^{\pm \pm} \to \mu^{\pm} \tau^{\pm} / \mu^{\pm} \mu^{\pm} / \tau^{\pm} \tau^{\pm} $, as they offer relatively large branching ratios. 
 
\begin{figure}[b]
	
	\includegraphics[width=8.5cm,height=7cm]{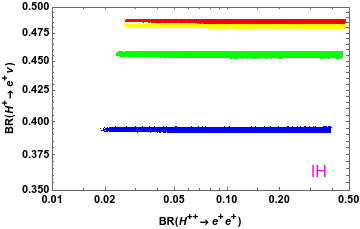}%
	\includegraphics[width=8.5cm,height=7cm]{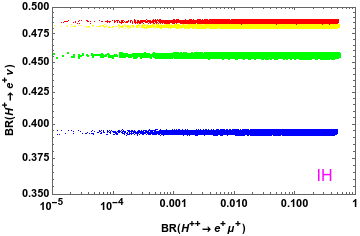}
	\includegraphics[width=12cm,height=7.5cm]{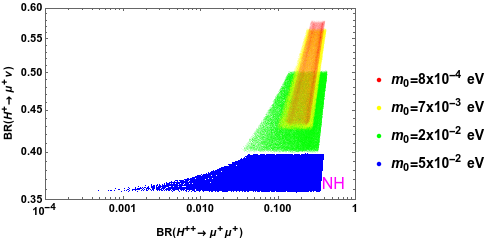}
	\caption{Variation of leptonic branching ratios of singly charged Higgs with the leptonic branching ratios of doubly charged Higgs. Here $\nu$ implies all the neutrino states $\nu_{1,2,3}$. See the texts for the details.} 
	\label{fig:corel}
\end{figure}
\item \textbf{Relating }  $\boldsymbol{H^{\pm\pm}}$ \textbf{and} $\boldsymbol{H^{\pm}}$ \textbf{Decays} 

\hspace{0.5cm}The doubly charged Higgs $H^{\pm \pm}$ as well as singly charged Higgs $H^{\pm}$ interact with the leptons through the same Yukawa couplings, that determine light neutrino masses. Therefore, the branching ratios of $H^{\pm \pm}$ into $l^{\pm} l^{\pm}$, and the branching ratio of $H^{\pm}$ into  $ l^{\pm} \nu$ are related. Fig.~\ref{fig:corel} shows the variation  of $H^{\pm} $ branching with   $ H^{\pm\pm} $ branching  for different leptonic decay channels.  Here we consider four illustrative samples  of lightest neutrino masses $(m_{0}= 0.0008,0.007,0.02,0.05 \ \text{eV})$ that covers almost entire allowed light neutrino spectrum. Different colour codes indicate different  values of $m_0$. The spread  of these color bands along horizontal and vertical direction represent the uncertainty in BR($H^{\pm\pm}\to l_{i}^{\pm}l_{j}^{\pm}$) and  BR($H^{\pm}\to l_{i}^{\pm}\nu$),  respectively. In  the upper left panel of Fig.~\ref{fig:corel},  we show the variation of BR($H^{\pm\pm}\to e^{\pm}e^{\pm}$) with the variation of  BR($H^{\pm}\to e^{\pm}{\nu}$), where we  assume  IH for neutrino mass ordering. There is a small variation in  BR($H^{\pm}\to e^{\pm}{\nu}$)  for a given value of  BR($H^{\pm\pm}\to e^{\pm}e^{\pm}$) that occurs due to the variation of oscillation parameters. Upper right panel of Fig.~\ref{fig:corel} represents the  variation of  BR($H^{\pm}\to e^{\pm}\nu$) with  BR($H^{\pm\pm}\to e^{\pm}\mu^{\pm}$), again  assuming IH as  neutrino mass ordering.  This  also shows similar features as the previous plot.  The plot in the lower panel  in Fig.~\ref{fig:corel} shows large variation of    BR($H^{\pm}\to \mu^{\pm}\nu$)  with  BR($H^{\pm\pm}\to \mu^{\pm}\mu^{\pm}$) for NH. For smaller $m_0$, the  BR($H^{\pm\pm}\to \mu^{\pm}\nu$) has a large dependency on oscillation parameters. 

 \end{itemize}
As we quantify  in the next section, the uncertainty in branching ratios can have large impact on the theory cross-section. 
\section{ Pair-production cross-section for $\sqrt{s}=13\, \text{TeV}$ LHC  \label{limit}}

\begin{figure}[h]
	\begin{subfigure}{0.5\textwidth}	
		\includegraphics[width=7.6cm,height=5cm]{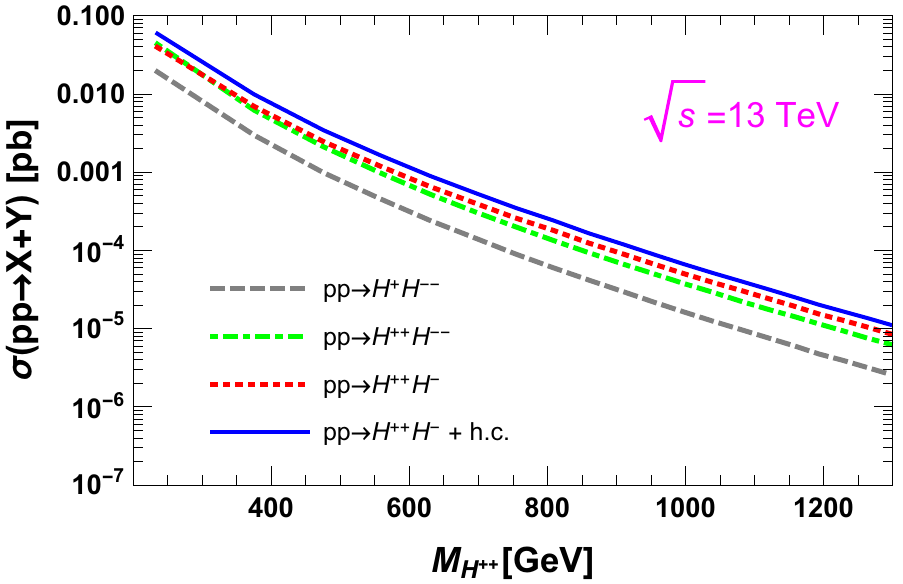}
	%	\caption{$\sqrt{s} = 13 $TeV}
		\label{fig:sfig1}
	\end{subfigure}%
	\begin{subfigure}{.5\textwidth}
		\includegraphics[width=7.6cm,height=5cm]{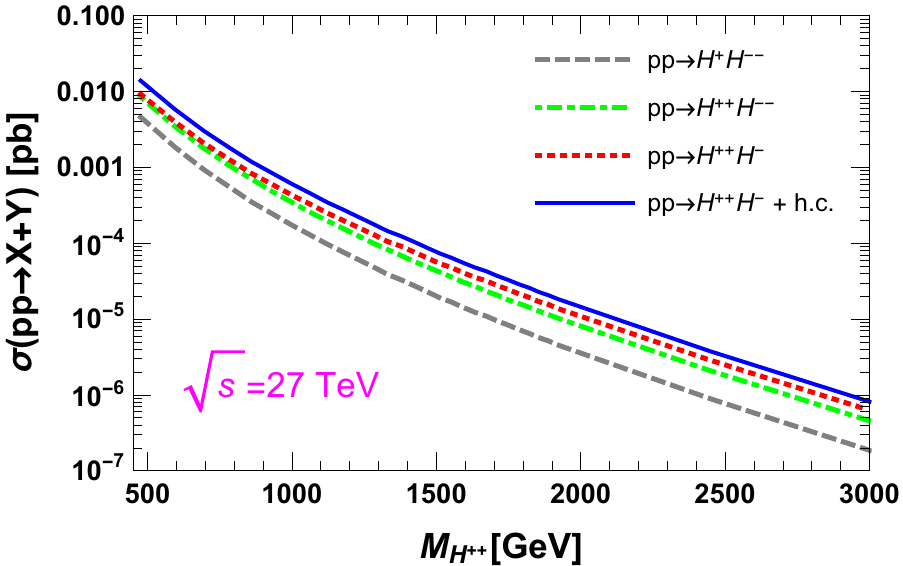}
		%\caption{$\sqrt{s} = 27 $TeV}
		\label{fig:sfig2}
	\end{subfigure}
	\caption{Pair and associated production cross-section of $H^{\pm\pm}$ as a function of $M_{H^{\pm\pm}}$.}
	\label{fig:cross}
\end{figure}

In Fig.~\ref{fig:cross}, we plot the  production cross-section of  ${H^{\pm\pm}}$ as a function of  $M_{H^{\pm\pm}}$ at LHC with $\sqrt{s}$ = 13 TeV. we also show the cross-section for a future $pp$ collider HE-LHC that can operate with center of mass energy 27 TeV. Here we show both pair ( $p p \to H^{++} H^{--}$)  and associated  ($p p \to H^{++} H^{-} + h.c.$)  production modes. Cross sections for the production of $ H^{++} H^{--}$(mediated by $\gamma^{\star}/Z^{\star}$) and  $ H^{++} H^{-}$(mediated by $W^{+\star}$) are comparable. As shown in Fig.~\ref{fig:cross}, the production cross-section of $ H^{+} H^{--}$(mediated by $W^{-\star}$) is smaller than that of $ H^{++} H^{-}$, which can be understood from parton distribution functions of proton. At $p\bar{p}$ collider both are the same. We consider a $K$-factor as  1.25 \cite{Muhlleitner:2003me} for the left panel of Fig.~\ref{fig:cross}. In our analysis we assume degenerate mass spectrum for the singly and doubly charged Higgs. 

 CMS and ATLAS collaboration have already placed constraint on  $M_{H^{\pm\pm}}$ by analyzing the leptonic decay channels of $ H^{\pm\pm}$ \cite{CMS-PAS-HIG-16-036, Aaboud:2017qph}. {A  degenerate mass spectrum for charged scalars  and   BR($ H^{\pm\pm} \to l_i^{\pm} l_j^{\pm} $)= 100\% have been assumed in the analysis. The  CMS analysis focussed on   the tri-lepton and four-lepton  final states originating from the leptonic decays of $ H^{\pm\pm} {\rm{and} } \, H^{\pm}$.} ATLAS searches considered   pair production of  $ H^{\pm\pm}$  and their subsequent decay into 
 $e^{\pm} e^{\pm} ,  e^{\pm} \mu^{\pm} ,  \mu^{\pm} \mu^{\pm}$  states. As a result of these searches, limits on $M_{H^{\pm\pm}}$ 
 vary between 770 GeV and 870 GeV at 95$\%$ C.L. CMS collaboration studied  both pair and associated production channels of $ H^{\pm\pm}$ and subsequent decay of $ H^{\pm\pm}$ and  $ H^{\pm}$  to different leptonic states. Limits on $M_{H^{\pm \pm}}$ obtained from the  combined study of both the channels  vary between 535 to  820 GeV  at 95$\%$ C.L,  for $100$\% branching to each leptonic state. This limit vary between 396 to 712 GeV, if only the  pair production channel is considered. The most stringent constraint $M_{H^{\pm\pm}} > 820 $ GeV  has been given by assuming $ H^{\pm\pm} \to e^{\pm} \mu^{\pm} $ decay, and this takes into account both 
 pair and associated productions. The CMS analysis~\cite{CMS-PAS-HIG-16-036}  has further considered  few benchmark points, and has given limits on $M_{H^{\pm \pm}}$. However, the PMNS mixing angle $\theta_{13}$ has been assumed as zero, that is  inconsistent with the present neutrino oscillation data.
  The above mentioned  searches  include  pair and associated production of $ H^{\pm\pm} $ and only its leptonic decay modes, so the observed limit on   $ M_{H^{\pm\pm}} $ is valid only for low triplet vev $v_\Delta \leq 10^{-4}$ GeV, where the di-leptonic branching is maximum. {As this is  evident from the discussion presented in the previous section,  the  maximum possible branching in each channel  can never be  $100\%$, rather can be at most $73\%$ (for $H^{\pm\pm} \to \mu^{\pm} \tau^{\pm}$ in NH).  Instead of considering  BR($H^{\pm\pm} \to l_i^{\pm} l_j^{\pm}$ ) = 100\%, we re-scale the theory cross-section with  appropriate branching ratios.  
 This somewhat  weakens  the individual bounds from different channels.
 In context of BNT model~\cite{Ghosh:2018drw} it has been shown that taking into account neutrino oscillation data one can lower the current CMS bound on $M_{H^{\pm\pm}} $.
   For illustration, we focus on the final states with $e^{\pm} e^{\pm} e^{\mp} e^{\mp}$, $e^{\pm} \tau^{\pm} e^{\mp} \tau^{\mp}$ and $e^{\pm} \mu^{\pm} e^{\mp} \mu^{\mp}$. Due to the absence of any cancellation region, the first channel  is the  least uncertain. We note that,  apart from the dependency on neutrino oscillation parameter,  the limit from individual channel  also depends on the value of lightest neutrino mass $m_0$.  }   
\begin{figure}[b]
	\includegraphics[scale=1,width=5.2cm,height=4.5cm]{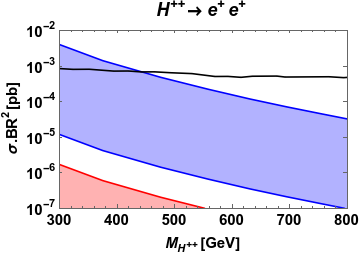}
	\includegraphics[scale=1,width=5.2cm,height=4.5cm]{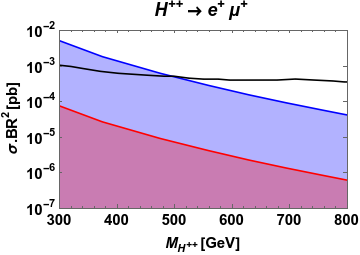}
	\includegraphics[scale=1,width=5.2cm,height=4.5cm]{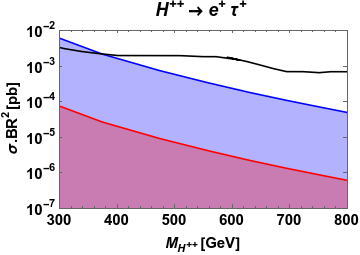}\\
	\includegraphics[scale=1,width=5.2cm,height=4.5cm]{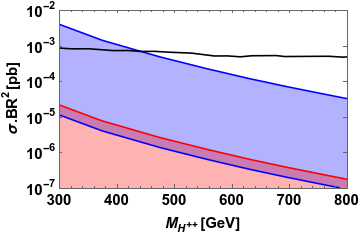}
	\includegraphics[scale=1,width=5.2cm,height=4.5cm]{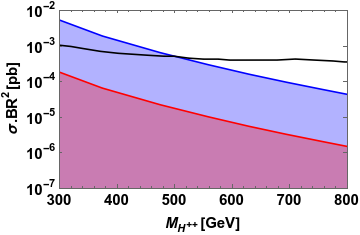}
	\includegraphics[scale=1,width=5.2cm,height=4.5cm]{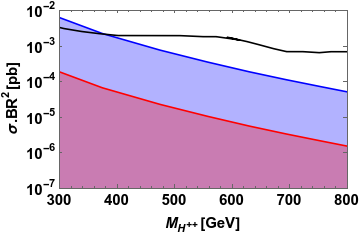}\\
	\includegraphics[scale=1,width=5.2cm,height=4.5cm]{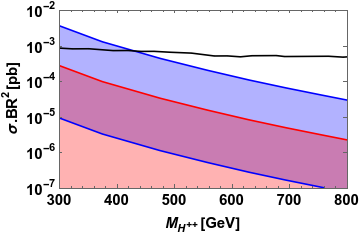}
	\includegraphics[scale=1,width=5.2cm,height=4.5cm]{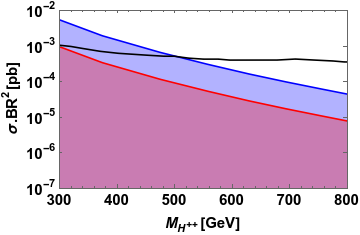}
	\includegraphics[scale=1,width=5.2cm,height=4.5cm]{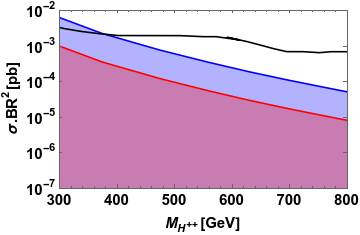}\\
	
	\caption{The blue (red) bands  for IH (NH)  correspond to the theory cross-section for the channel $p p \to H^{++} H^{--} \to l_i^{+} l_j^{+} l_k^{-} l_l^{-} $  obtained by including  $3\sigma$ variation  of neutrino oscillation parameters. Black line represents  the observed limit from  CMS  analysis~\cite{CMS-PAS-HIG-16-036}. The horizontal panels in  row 1-3 represent $m_0=(0.0008,0.007,0.02) \  \text{eV}$. In 1st, 2nd and 3rd columns we consider  the decay of $ H^{\pm\pm} $  to $e^{\pm} e^{\pm} $ , $e^{\pm} \mu^{\pm} $ and $e^{\pm} \tau^{\pm} $, respectively.}
	\label{fig:band}
\end{figure}

{In Fig.~\ref{fig:band},  we show  the production cross-section of  $p p \to H^{++} H^{--} \to e^{+} e^{+} e^{-} e^{-}$, $e^{+} \mu^{+} e^{-} \mu^{-}$, and  $e^{+} \tau^{+} e^{-} \tau^{-}$   at LHC for $\sqrt{s} = 13\,  $TeV. The coloured band represents the variation of cross-section due to $3\sigma$ uncertainty in neutrino oscillation parameters. As illustrative points, we choose three values of lightest neutrino masses  $m_{0}={0.0008, 0.007,0.02} \ \text{eV}$, that falls in hierarchical  mass regime. The blue (red) band corresponds to IH (NH)  neutrino mass spectrum. The 
black line represents  the observed limit from   13 TeV CMS analysis \cite{CMS-PAS-HIG-16-036}. For a given value of the lightest neutrino mass $m_0$, the upper boundary in these bands  is determined from   $\sigma(p p \to H^{++} H^{--})$ folded with  the square of maximum possible  BR($H^{\pm\pm} \to l_i^{\pm} l_j^{\pm}$). Similarly, the  lower line represents the  minimum value of  BR($H^{\pm\pm} \to l_i^{\pm} l_j^{\pm}$).  Couple of points are in order :

\begin{itemize}
\item
The total cross-section has a large variation, specially for  $e^+\mu^+e^-\mu^-$ and   $e^{+} \tau^{+} e^{-} \tau^{-}$ channels. The  $e^{+}e^{+} e^{-}e^{-}$ channel in IH is the least uncertain, as this has a definite lower value of the cross-section. 
\item
Due to relatively smaller  branching ratio,  the cross-section in NH for these modes are lower than the maximal possible cross section in IH. 
\item
The drop in cross-section for $ e^{+} \mu^{+} e^{-} \mu^{-}$ and  $e^{+} \tau^{+} e^{-} \tau^{-}$ occurs, due to the cancellation between different terms in $M_{12}^{\nu}$ and $M_{13}^{\nu}$. 
\end{itemize}

Taking into account the  branching ratios, the limit from each of the leptonic channels  somewhat weakens, as compare to the analysis presented in \cite{CMS-PAS-HIG-16-036}. However, the combined limit might be comparable to that analysis. For the above modes, IH can give the best constraint. The cross-section for NH is order of magnitude smaller in the hierarchical limit, and therefore, competitive limits  can not be placed on $M_{H^{\pm \pm}}$ in the above channels, if light neutrinos follow NH. We tabulate the predicted value of maximum possible branching ratios in  Table.~\ref{fig:limit}, where each entry represents the maximum possible value of BR($ H^{\pm} \to l_i^{\pm} l_j^{\pm} $) for a given value of  $m_0$. The value within the bracket denotes  the best lower limit on $M_{H^{\pm\pm}}$, from each channel. 

 In Fig.~\ref{masslimit_NH}, we present plots similar to that of Fig.~\ref{fig:band}, considering the decay of $H^{\pm\pm}$  to $\mu^{\pm} \mu^{\pm} $ , $\mu^{\pm} \tau^{\pm} $ and $\tau^{\pm} \tau^{\pm} $. The different plots represent the production cross-section of  $p p \to H^{++} H^{--} \to \mu^{+} \mu^{+} \mu^{-} \mu^{-}$, $\mu^{+} \tau^{+} \mu^{-} \tau^{-}$ and $\tau^{+} \tau^{+}\tau^{-} \tau^{-}$ at LHC for $\sqrt{s} = 13\, $ TeV. We use the same color code and the same values of lightest neutrino masses as used in Fig.~\ref{fig:band}. For these modes, the maximal possible cross-section in NH are higher as compared to IH. Therefore, competitive limits on $M_{H^{\pm\pm}}$ can be derived in the above mentioned channels, if NH is assumed. The derived limits on $M_{H^{\pm\pm}}$ are tabulated in Table.~\ref{fig:limit}.
\begin{figure}[h]
	\includegraphics[scale=1,width=5.2cm,height=4.3cm]{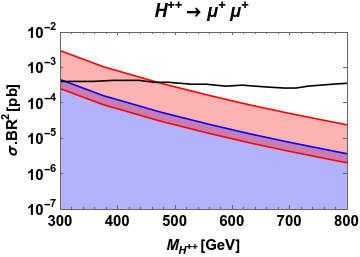}
	\includegraphics[scale=1,width=5.2cm,height=4.3cm]{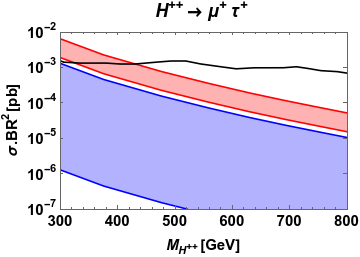}
	\includegraphics[scale=1,width=5.2cm,height=4.3cm]{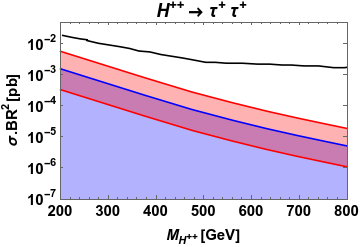}\\
	\includegraphics[scale=1,width=5.2cm,height=4.3cm]{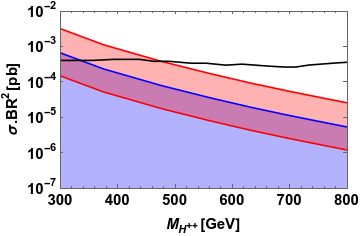}
	\includegraphics[scale=1,width=5.2cm,height=4.3cm]{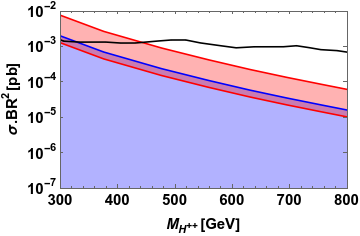}
	\includegraphics[scale=1,width=5.2cm,height=4.3cm]{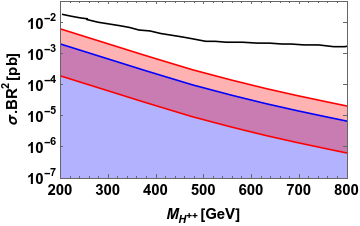}\\
	\includegraphics[scale=1,width=5.2cm,height=4.3cm]{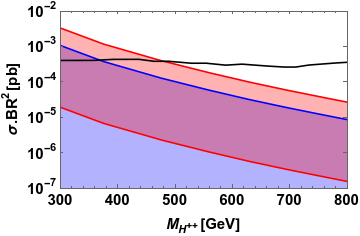}
	\includegraphics[scale=1,width=5.2cm,height=4.3cm]{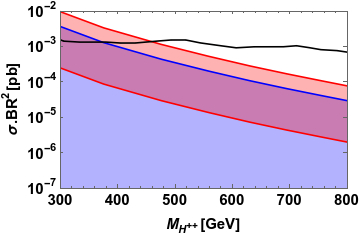}
	\includegraphics[scale=1,width=5.2cm,height=4.3cm]{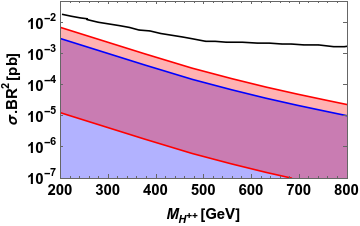}\\
	
	\caption{The blue (red) bands  for IH (NH)  correspond to the theory cross-section for the channel $p p \to H^{++} H^{--} \to l_i^{+} l_j^{+} l_k^{-} l_l^{-} $ obtained by including  $3\sigma$ variation  of neutrino oscillation parameters. Black line represents  the observed limit from  CMS analysis~\cite{CMS-PAS-HIG-16-036}. The horizontal panels in  row 1-3 represent $m_0=(0.0008,0.007,0.02) \  \text{eV}$. In 1st, 2nd and 3rd columns we consider  the decay of $ H^{\pm\pm} $  to $\mu^{\pm} \mu^{\pm} $ , $\mu^{\pm} \tau^{\pm} $ and $\tau^{\pm} \tau^{\pm} $, respectively.}
	\label{masslimit_NH}
	
\end{figure}	

In CMS search combined limits have been presented which result from  the combined analysis  of both the pair and associated production channels. Also, for the benchmark studies, their analysis combines different leptonic modes. Such a study for the combined limit is beyond the scope of this paper. 
    \begin{table}[h]
    	\begin{tabular}{|c|c|c|c|c|}
    		\hline
    		\multicolumn{5}{|c|}{Maximum value of BR($ H^{\pm\pm} \to l_i^{\pm} l_j^{\pm}$) $(M_{H^{\pm\pm}} \text{[GeV]})$}            \\ \hline
    		Decay mode &$m_{0}=0.0008$ eV  & $m_{0}=0.007$ eV \quad & $m_{0}=0.02$ eV \qquad & Mass Hierarchy \\ \hline
    		$e^{\pm}e^{\pm}$ & 0.478 (435)&0.476 (435) &0.454 (424)  & IH  \\ \hline
    		$e^{\pm}\mu^{\pm}$&  0.537 (495)&0.547 (503)&  0.552 (503)  &IH   \\ \hline
    		$e^{\pm}\tau^{\pm}$ & 0.583 (373)&0.594 (376)&  0.594 (376) & IH   \\ \hline
    		$\mu^{\pm}\mu^{\pm}$ &0.410 (465)  &0.424 (478)&0.434 (482)&NH\\ \hline
    		$\mu^{\pm} \tau^{\pm}$ &0.604 (428) & 0.656 (440) &  0.735 (450) &  NH  \\ \hline
    		$ \tau^{\pm}\tau^{\pm}$ & 0.363($<200$)&0.382($<200$)&0.404($<200$) & NH  \\ \hline
    	\end{tabular}
    	\caption{ Maximum possible branching ratio for the decay mode $ H^{\pm\pm} \to l_i^{\pm} l_j^{\pm} $. We also show the corresponding lower limit on $M_{H^{\pm\pm}}$ in bracket obtained from the channel $p p \to H^{++} H^{--} \to l_i^{+} l_j^{+} l_i^{-} l_j^{-}$ (Here $l_i^+ = e^+ / \mu^+ / \tau^+ $). We use \cite{CMS-PAS-HIG-16-036} to derive the limits.}
    	\label{fig:limit}
    \end{table}

 As discussed in \cite{Dev:2016dja,Borah:2018yxd}, the $3\sigma$ sensitivity reach of a doubly-charged Higgs in minimal left  right symmetric model is less than 1.3 TeV, where a c.m.energy $\sqrt{s}=14$ TeV and $3\ \text{ab}^{-1}$  integrated luminosity have been assumed. Another study on charged Higgs in the context of minimal super-symmetric model \cite{Aboubrahim:2018tpf} compares the discovery potential of HL-LHC and HE-LHC. In Ref.~\cite{Crivellin:2018ahj}, the phenomenology of a $SU(2)$-singlet doubly charged scalar has been discussed at HL-LHC.
 In Ref.~\cite{Mitra:2016wpr}, pair production of $H^{\pm\pm}$ and its subsequent decay to same-sign dilepton has been explored at 13 TeV and 14 TeV LHC for $M_{H^{\pm\pm}}$ upto 700 GeV. Fig.~6 of Ref.~\cite{Mitra:2016wpr} shows that 13 TeV and 14 TeV cross-sections are not very different.
 It is evident from Fig.~\ref{fig:cross}, that the  pair production cross-section of $H^{\pm\pm}$ at 13 TeV LHC becomes smaller for higher mass. For $M_{H^{\pm\pm}}=1.3$ TeV, the cross-section is less than $10^{-5}$ pb, whereas at HE-LHC it is around $10^{-4}$ pb. Therefore, assuming $3\ \text{ab}^{-1}$ ($15\ \text{ab}^{-1}$) integrated luminosity, approximately 60 (3000) number of $H^{\pm\pm}$ with mass $1.3$ TeV can be produced at HL-LHC (HE-LHC). Increase in sensitivity for higher mass range will be possible for HE-LHC. Therefore, to study the discovery prospects of heavier  $H^{\pm \pm}$ and $ H^{\pm}$, we consider higher center of mass energy, i.e., the  HE-LHC setup  with $\sqrt{s}=27$ TeV \footnote{While preparing the manuscript for this work, Ref.~\cite{deMelo:2019asm} appeared in arXiv. This considers pair-production of doubly charged Higgs and subsequent decays at HE-LHC. We consider both pair and associated production, oscillation parameter dependency, and the correct values of the leptonic branching ratios which has largely been overlooked in the literature.}.

 In the next section we present the collider analysis for multilepton signatures of $H^{\pm\pm}$, where we assume  BR($ H^{\pm\pm} \to e^{\pm} \mu^{\pm} $) = 0.547, corresponding to $m_0 = $0.007 eV,  and IH as neutrino mass ordering. The other branching ratios are given in Table.~\ref{fig:BP1}. For completeness,  in our analysis, we however consider all  leptonic modes, with their corresponding  branching ratios. Note that, other than the pair-production by Drell-Yan process, the photon fusion can also contribute to the pair-production of doubly charged Higgs. It has been pointed out in~\cite{Cai:2017mow}, that for 13 TeV, the channel contributes at most 10$\%$ to the pair-production of doubly charged Higgs states. However, there are different issues, regarding  large uncertainties in  PDFs. Therefore one needs to evaluate this channel carefully. We do not consider this channel in our present analysis.

\section{Multi-lepton Signals from $H^{\pm \pm}$ and $H^{\pm}$ for  $\sqrt{s} = 27\, \text{TeV}$ HE-LHC \label{multilep}}
We consider the set-up for a future $p p $ collider HE-LHC, that can operate with  a c.m.energy $\sqrt{s}=27$ TeV, and analyse the tri-lepton and four-lepton channels in detail. To simulate the signal samples, we   implement the model in 
FeynRules(v2.3) \cite{Alloul:2013bka}. The  UFO output   is then fed  into  MadGraph5\_aMC@NLO(v2.6) \cite{Alwall:2014hca}   that  generates  the parton-level  events. We use the default PDFs NNPDF23LO1 \cite{Ball:2013hta} for computation.  We perform parton showering and hadronization  with Pythia8 \cite{Sjdostrand:2007gs} and carry out the detector simulations  with Delphes(v3.4.1) \cite{deFavereau:2013fsa}. Finally data analysis and ploting is done in ROOT(v6.14/04) \cite{Brun:1997pa}. 
We choose the degenerate mass spectrum for charged Higgses for which the most promising signals are $4$ lepton and $3$  lepton final states,  arising  from pair  and associated production of doubly charged Higgs.
\subsection{$4l$ Final State }
{This originates from the pair-production of $H^{++}$ and its subsequent decays  $H^{\pm \pm} \to l^{\pm} l^{\pm}$. Therefore, the signal is represented by the following chain, }
\begin{itemize}
\item
Signal : $p p \to H^{++} H^{--} \to l_i^{+} l_j^{+} l_k^{-} l_l^{-} $, ( with  $l_i^{\pm} = e^{\pm} / \mu^{\pm} / \tau^{\pm} $).
\end{itemize}
The $\tau$ in the final state,  further decays into fully hadronic, or leptonic final states. For our analysis,  we consider leptonic decays of $\tau$, and therefore, collect all the event samples with $e, \mu$ in the final state. There are a number of SM processes that can mimic 
the signal, hence are considered as SM background. Here we list the following processes as dominant SM background:
\begin{itemize}
\item $ ZZ\ :\ p p \to Z Z \to  l_i^{+} l_i^{-} l_j^{+} l_j^{-} $
\item $t \bar{t}  Z \ : \ p p \to t \bar{t}  Z \to l_i^{+} l_j^{-} l_k^{+} l_k^{-} + b \bar{b} + \slashed{E}_{T}$
\item $t \bar{t}  W \ : \ p p \to t \bar{t}  W^{\pm} \to l_i^{+} l_j^{-} l_k^{\pm} + b \bar{b} + \slashed{E}_{T}$
\item $t \bar{t} \ : \  p p \to t \bar{t} \to l_i^{+} l_j^{-} + b \bar{b} + \slashed{E}_{T} $
\item $VVV \ ( V = Z \ \text{or} \ W^\pm) \ : \  p p \to VVV ,\  V \to l_i^+ l_i^- \ 
\text{or} \ l_i^\pm \slashed{E}_{T} \text{or} \ jj $
\item $WZ \ :\ p p \to W^{\pm} Z \to  l_i^{\pm} l_j^{+} l_j^{-} + \slashed{E}_{T}$.
\item $t \bar{t}t \bar{t}  \ : \ p p \to t \bar{t} t \bar{t} \to l_i^{+} l_j^{-} l_k^{+} l_m^{-} + 2b \bar{b}  + \slashed{E}_{T}$
\end{itemize}

Among all these  backgrounds,  $ZZ, \ t \bar{t}  Z$, $W^{\pm} W^{\mp} Z$  processes  lead to irreducible backgrounds. However, a few  other SM processes, such as, $t \bar{t},\ t \bar{t}W, \ W Z$   with their subsequent decays can also give rise to four-lepton final states, due to the misidentification of jets as leptons.   Multi-lepton events ($N^{l} > 4$) from $p p \to ZZZ \to 6 l$ ($l=e, \mu$)  can also mimic the signal due to detector inefficiency in lepton reconstruction, or if the lepton is too soft, and does not pass the selection cuts. Additionally,   one of the $Z$ bosons in the above mentioned background can  decay to two hadronic taus, that can also mimic the signal. As we will show below, most of the backgrounds are reduced  significantly after imposing $Z$-veto, as well as, selecting a window on the $l^{\pm} l^{\pm}$ invariant mass.  

In  Fig.~\ref{fig:cross}, we show the pair-production cross-section  of $H^{\pm \pm}$ for  $\sqrt{s} \ = 27$ TeV. The cross-section  varies from $10^{-2}$ pb for $M_{H^{ \pm \pm}} = 400$ GeV
 to  $10^{-6}$  pb for $M_{H^{ \pm \pm}} = 3$ TeV. As the cross-section is gradually decreasing with increasing mass, it will be difficult to probe very heavy $H^{\pm \pm}$. 
 Here we present a benchmark point with $M_{H^{\pm \pm}}$ = 1 TeV to show a detail cut-efficiency. We consider a triplet vev   $v_{\Delta}$  as $10^{-8}$ GeV. We re-iterate that,  for the analysis, we consider IH  neutrino mass ordering and  the following set of oscillation parameters, for which $H^{\pm \pm } \to e^{\pm}\mu^{\pm}$  is the most dominant decay channel with a branching ratio  0.547:  

\begin{itemize}
\item $\theta_{12} = 0.6567, \ \theta_{13} = 0.1567, \  \theta_{23} = 0.7385,$
\item $ \phi_1 = 3.0614, \ \phi_2 = 5.9, \ \delta = 0.2029,$ 
\item  $m_1 = 0.04902$ eV, $m_2=0.04973$ eV and $m_3=0.007$ eV.
\end{itemize} 
This set of parameters is assumed because it puts the strongest limit on $M_{H^{\pm\pm}}$, as evident from Table.~\ref{fig:limit}.  Another reason for selecting this particular set of parameters is to reduce the value of  BR($H^{\pm\pm}\to \tau^{\pm} \tau^{\pm}$).
Branching ratio of $ H^{\pm\pm}$ decays to different leptonic flavour states for this set of parameter are shown in Table~\ref{fig:BP1}. {Note that, the doubly charged Higgs pre-dominantly decays to $e^\pm\mu^\pm$ and $e^\pm\tau^\pm$ final state.}
\begin{table}[t]
	
	\begin{tabular}{|c|c|c|c|c|c|c|}
		\hline	
		& $e^\pm e^\pm$&$e^\pm\mu^\pm$ &$e^\pm\tau^\pm$&$\mu^\pm \tau^\pm$   & $\mu^\pm \mu^\pm$ &$\tau^\pm \tau^\pm$ \\ \hline
		BR&0.026 &0.547&0.365  &0.001 &0.001&0.053 \\ \hline
	\end{tabular}
	\caption{{Branching ratio of $H^{\pm \pm}$ into different leptonic states in case of IH  for $m_0 = 0.007$ eV.}}
	\label{fig:BP1}
\end{table}

{We apply the following set of basic cuts on transverse momentum, pseudo-rapidity, and separation between two leptons, } $P_{T}(l) > 10$ GeV  ($l = e \ \text{or} \ \mu$), $\abs{\eta(l)} < 2.5$, $\Delta R_{ll} > 0.4$,  at the time of event generation in MadGraph5. For  detector level analysis, the isolation condition for a lepton($e,\mu$) is defined as: $\frac{\sum P_T(x)}{P_T(l)}<0.2$, where $\sum P_T(x)$ is the scalar sum of transverse momenta of all particles with in a cone of radius $\Delta R<0.4$ around the lepton direction and $P_T(l)$ is the transverse momentum of lepton.  
We assume a jet misidentification rate\footnote{ It is defined as the rate by which jets are identified
	as leptons. There is a small chance for low $P_T$ jets to be identified as leptons.  Although it is small, a significant background cross-section can be resulted from misidentification  because of the large production cross section of QCD jets at the LHC.} of $10^{-3}$~\cite{Sirunyan:2018fpa}. 
\begin{figure}[b]
	\includegraphics[width=8.6cm,height=5.5cm]{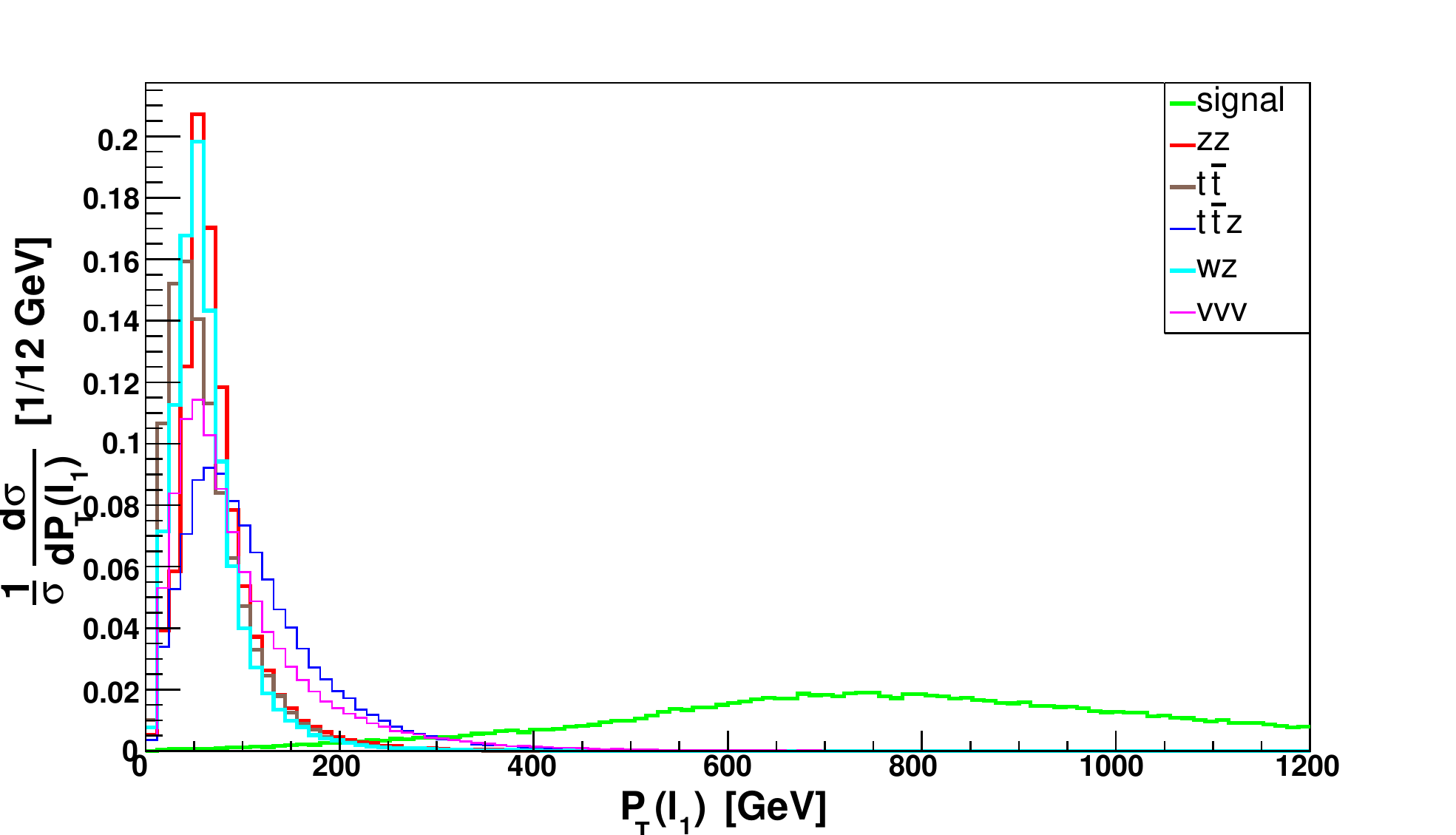}%
	\includegraphics[width=8.6cm,height=5.5cm]{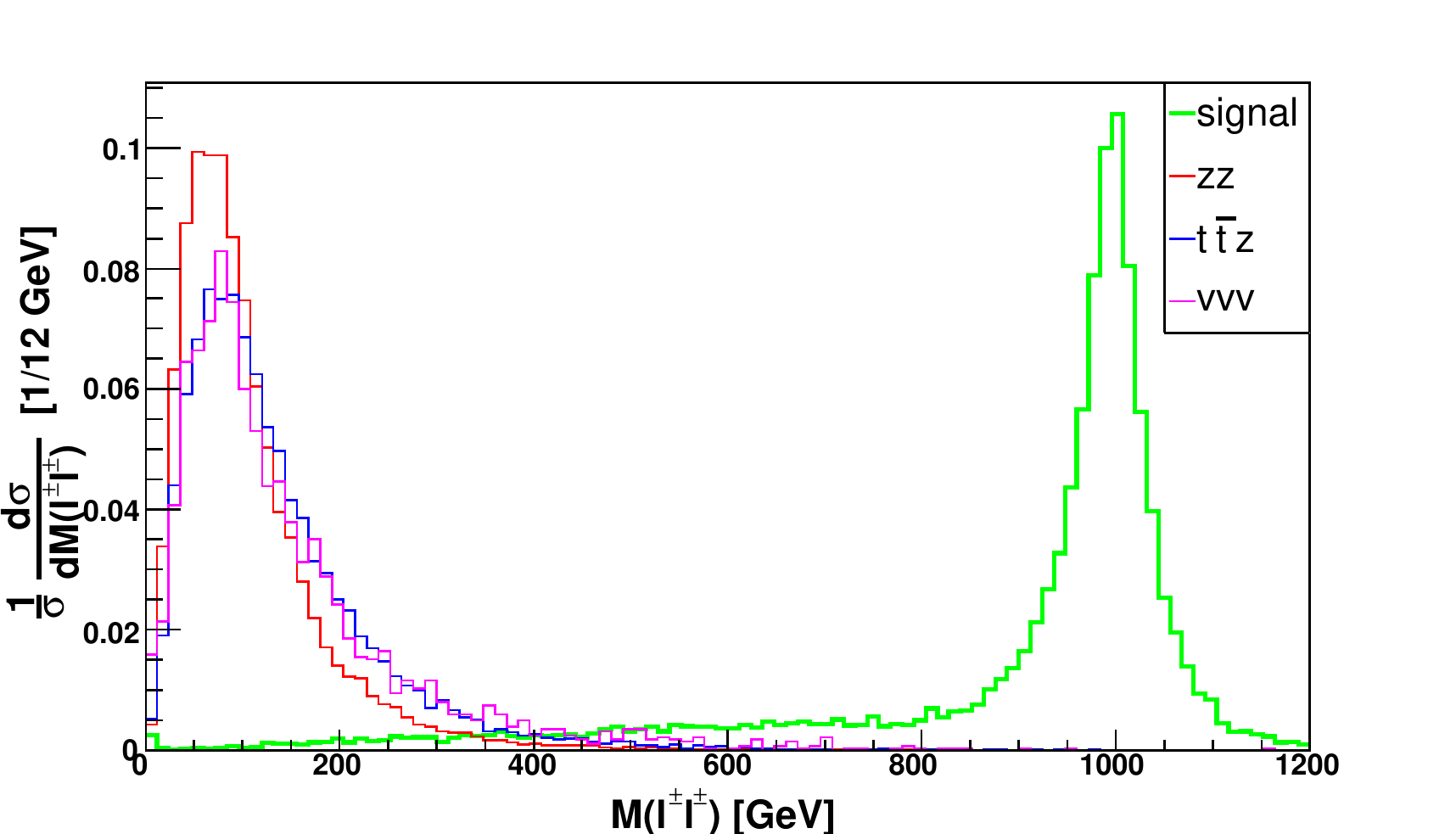}
	\includegraphics[width=9cm,height=5.5cm]{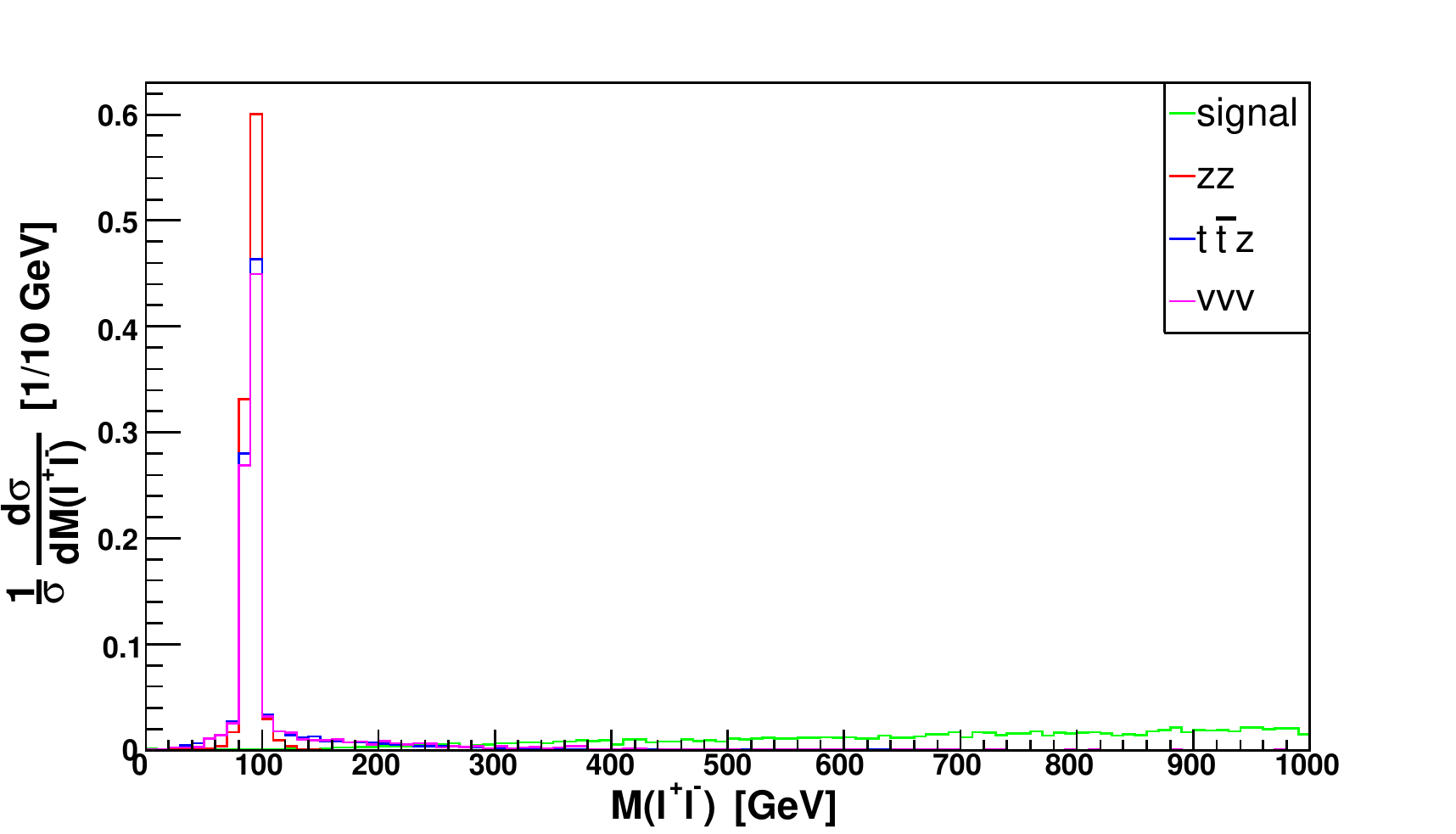}
	\caption{Normalised distributions of transverse momentum of leading lepton $P_T(l_1)$, same-sign di-lepton invariant mass $M(l^{\pm}l^{\pm})$, opposite sign di-lepton invariant mass $M(l^+l^-)$ for both $4l$ signal and background events.}
	\label{fig:dist}
\end{figure}

 In Fig.~\ref{fig:dist}, we plot distribution of  different kinematical variables for both signal and SM backgrounds. Top left plot in Fig.~\ref{fig:dist}  shows the transverse momentum distribution of leading lepton. It is evident  from this figure that most of the background lies in low $P_{T}(l_1)$ region and the signal peaks at high $P_{T}(l_1)$ for high $M_{H^{\pm \pm}}$. Top right plot in Fig.~\ref{fig:dist}, shows the distribution of  same-sign di-lepton invariant mass $M(l^\pm l^\pm)$. Here the signal distribution  peaks at  $M_{H^{\pm \pm}}$, which is very clear and  well separated from backgrounds. Such a distinguished peak of $M_{H^{\pm \pm}}$  at a high value of $M(l^{\pm} l^{\pm})$  distribution helps to discover $H^{\pm\pm}$.  
In the lower panel of Fig.~\ref{fig:dist}, we show the  distribution of opposite sign di-lepton invariant mass, which shows that most of the dominant backgrounds peak around $Z$ boson mass. Therefore, a veto on opposite sign di-lepton invariant mass around $M_Z$ will reduce most of the backgrounds involving  $Z$.
 
 {The  above distributions motivate  to  consider  the following set of selection cuts that suppress  backgrounds:}
\begin{itemize}	
\item $ \rm{A_1}$ : $ \ N_{l} = 4$. we demand 4 isolated leptons in the final state. The leptons are $e,\mu$.
 \item $ \rm{A_2}$ : We demand sum of charges of four leptons to be zero.
 \item  $ \rm{A_3:}$$ \ \abs{M(l^+ l^-) - M_{Z}} > 10 $ GeV. To remove the backgrounds including atleast one $Z$ boson, we veto the lepton pairs with the same flavor but opposite charges inside the mass window $\abs{M(l^+ l^-) - M_{Z}} < 10$ GeV.
\item$\rm{A_4:}$$ \ \abs{M(l^\pm l^\pm) - M_{H^{\pm\pm}}} \leqslant 50$. 
 The  signal events are selected  by demanding a window on same-sign di-lepton invariant mass in between $M_{H^{\pm\pm}}\pm 50$ GeV.

\end{itemize}

In Table.~\ref{fig:cstable}, we show the changes in  signal and background cross-sections after  each  selection cut. 
\begin{itemize}
	\item $c_1 \ : \  N_{l} = 4$.
		\item $c_2 \ : \ c_1 \ \text{and} \ \Sigma l_{charge} = 0$ .
	\item $c_3 \ : \ c_2 \ \text{and} \ \abs{M(l^+ l^-) - M_{Z}} > 10$ GeV.
	\item $c_4 \ : \ c_3   \ \text{and} \ \abs{M(l^\pm l^\pm) - M_{H^{\pm\pm}}} \leqslant 50$.
\end{itemize}

From Table.~\ref{fig:cstable}, the signal cross-section before applying cut is  0.2683 fb for $M_{H^{\pm \pm}}=$ 1 TeV. Most dominant irreducible background appears from  $ZZ$  with a cross-section 83.559 fb. The channels
  $t \bar{t} \ \text{and} \ WZ  $ have huge cross-section compared to other backgrounds but they result in very less number of four lepton events.  Demanding  four leptons in the final state reduces the background cross-section to a significant extent. Invariant mass window on same-sign di-lepton finally help to suppress almost all background events. It should be noted that, we consider  $t\bar{t}$\ \text{and} $WZ$ backgrounds with 0-parton. In our analysis, particles of interest are not jets but leptons as the final state contains multilepton. We implimented stringent selection criteria on leptons. Hence we expect involvement of multi-parton processes will not subtantialy change the final results. However, we check the contribution of multi-parton $t\bar{t}$\ \text{and} $WZ$ processes, performing  MLM matching in MadGraph5\_aMC@NLO(v2.6)~\cite{Alwall:2014hca} and Pythia8~\cite{Sjdostrand:2007gs}. The result we obtain is similar to that in case of 0-parton backgrounds.

  \begin{table}[h]
  	
  	\begin{tabular}{|c|c|c|c|c|c|c|}
  		\hline
  		& \multicolumn{1}{c|}{	$\sigma$ [fb] for signal } & \multicolumn{5}{c|}{$\sigma$ [fb] for backgrounds} \\ \hline
  		& $M_{H^{\pm\pm}}$ = 1 TeV    	& $\ ZZ \ $ & $\ t \bar{t}\ $  & $t \bar{t}Z$ & $\ WZ \ $ & $VVV$       \\ \hline
  		before cut &  0.2683    & 83.559    &  142075   &  14.413   &  702.333  & 9.49                 \\ \hline
  		after  $c_1$ &  0.0597  &  18.56   &   9.9452  &   2.2616  &  0.9341  & 0.2668             \\ \hline
  		after $c_2$ & 0.0591& 18.5035    &  9.9453  &  2.2368   &  0.48461   &   0.2568      \\ \hline
  		after $c_3$    &    0.0589 &   0.2031 & 7.1037& 0.363    & 0.0913     & 0.0407     \\ \hline
  		after $c_4$ &  0.0194& $\approx 0$  & $\approx 0$ & $\approx 0$   & $\approx 0$     &   $\approx 0$       \\ \hline
  		
  	\end{tabular}
  	\caption{ Signal ( $p p \to H^{++} H^{--} \to l_i^{+} l_j^{+} l_k^{-} l_l^{-} $) and background cross-sections for $\sqrt{s}= 27$ TeV after the different  selection cuts for the channel. Here $l_i^+ = e^+ / \mu^+  $. }
  	\label{fig:cstable}
  \end{table}
  
 Note that, although we present the $p p \to ZZ \to 4l$ background in Table.~\ref{fig:cstable},  we also estimate  $p p \to 4l$, that  includes  virtual photon contribution.  The channel $pp \to l_i^{+} l_j^{+} l_k^{-} l_l^{-} $  has cross-section 117.1 fb.} We find a cross-section 2.8 fb after applying  cut $c_3$. However after cut $c_4$,  the  cross-section becomes negligibly small. {This is expected, as we choose a very large value of same-sign di-lepton invariant mass, for which this background already falls off. We check the contribution of $t \bar{t}t \bar{t}$ background, for which the cross-section is 0.276 fb. After applying cuts $c_2$ and $c_3$ the  cross-sections are 0.027 fb and 0.023 fb, respectively. After cut $c_4$  this cross-section is reduced to a insignificant value.
 In addition, we also check  $t \bar{t} W$ background, which  after cut  $c_4$ gives insignificant contribution. Although the SM background cross-section is much higher than that of signal before applying cut $c_1$,  the backgrounds  become insignificant  after applying  selection cuts. We find that, with 1000 $\rm{fb}^{-1}$ luminosity, 19  events  can be obtained for $M_{H^{\pm \pm}} = 1$ TeV. We give the variation of number of events versus mass of doubly charged Higgs in Fig.~\ref{fig:27_pred}.
 \subsection{$3l$ Final state}
  Here we consider   the signal containing tri-lepton (two same-sign lepton and other of opposite sign) and missing transverse energy $\slashed{E}_{T}$ in the final state. Associated production of $H^{\pm\pm}$ with $H^{\pm}$ and their subsequent leptonic decay dominantly contribute to the desired signal events. However pair production of $H^{\pm\pm}$ also contribute to the same when atleast one hadronically decaying tau lepton is present in the decay products of $H^{\pm\pm}$. Therefore,  the signal events we are analysing  originate  from the following decay chains:

 \begin{itemize}
\item PP : $p p \to H^{++} H^{--} \to l_i^{+} l_j^{+} l_k^{-} l_l^{-} $ (where $l_i^+ = e^+ / \mu^+ / \tau^+ $)
\item AP : $p p \to H^{++} H^{-} \ + h.c. \to l_i^{+} l_j^{+} l_k^{-} \nu$ (where $l_i^+ = e^+ / \mu^+ / \tau^+ $)
\end{itemize}
 We consider the following dominant SM backgrounds:
 \begin{itemize}
 	\item $WZ \ :\ p p \to W^{\pm} Z \to  l_i^{\pm} l_j^{+} l_j^{-} + \slashed{E}_{T}$.
 	\item $ ZZ\ :\ p p \to Z Z \to  l_i^{+} l_i^{-} l_j^{+} l_j^{-} $
 	\item $VVV \ ( V = Z \ \text{or} \ W^\pm) \ : \  p p \to VVV ,\  V \to l_i^+ l_i^-/  \ l_i^\pm \slashed{E}_{T}/  jj.$
 	\item $t \bar{t}  W \ : \ p p \to t \bar{t}  W^{\pm} \to l_i^{+} l_j^{-} l_k^{\pm} + b \bar{b} + \slashed{E}_{T}$
 	\item$DY \ :\ p p \to Z / \gamma \to l_i^+ l_i^-$
 	\item $t \bar{t} \ : \  p p \to t \bar{t} \to l_i^{+} l_j^{-} + b \bar{b} + \slashed{E}_{T} $
 	\item $t \bar{t}  Z \ : \ p p \to t \bar{t}  Z \to l_i^{+} l_i^{-} l_j^{+} l_j^{-} + b \bar{b} + \slashed{E}_{T}$
 \end{itemize}
$WZ$ and $VVV$ channels result in irreducible  background events with comparatively higher $\slashed{E}_{T}$. $ZZ$ and $t \bar{t}Z$ also contribute to the background when one among the leptons in the final state is a hadronic tau or is left undetected. Drell-Yan ($DY$), $ t \bar{t}$ process  give tri-lepton events when a jet  fakes as a lepton.\\
  \begin{figure}[b]
 	\includegraphics[width=9cm,height=5.5cm]{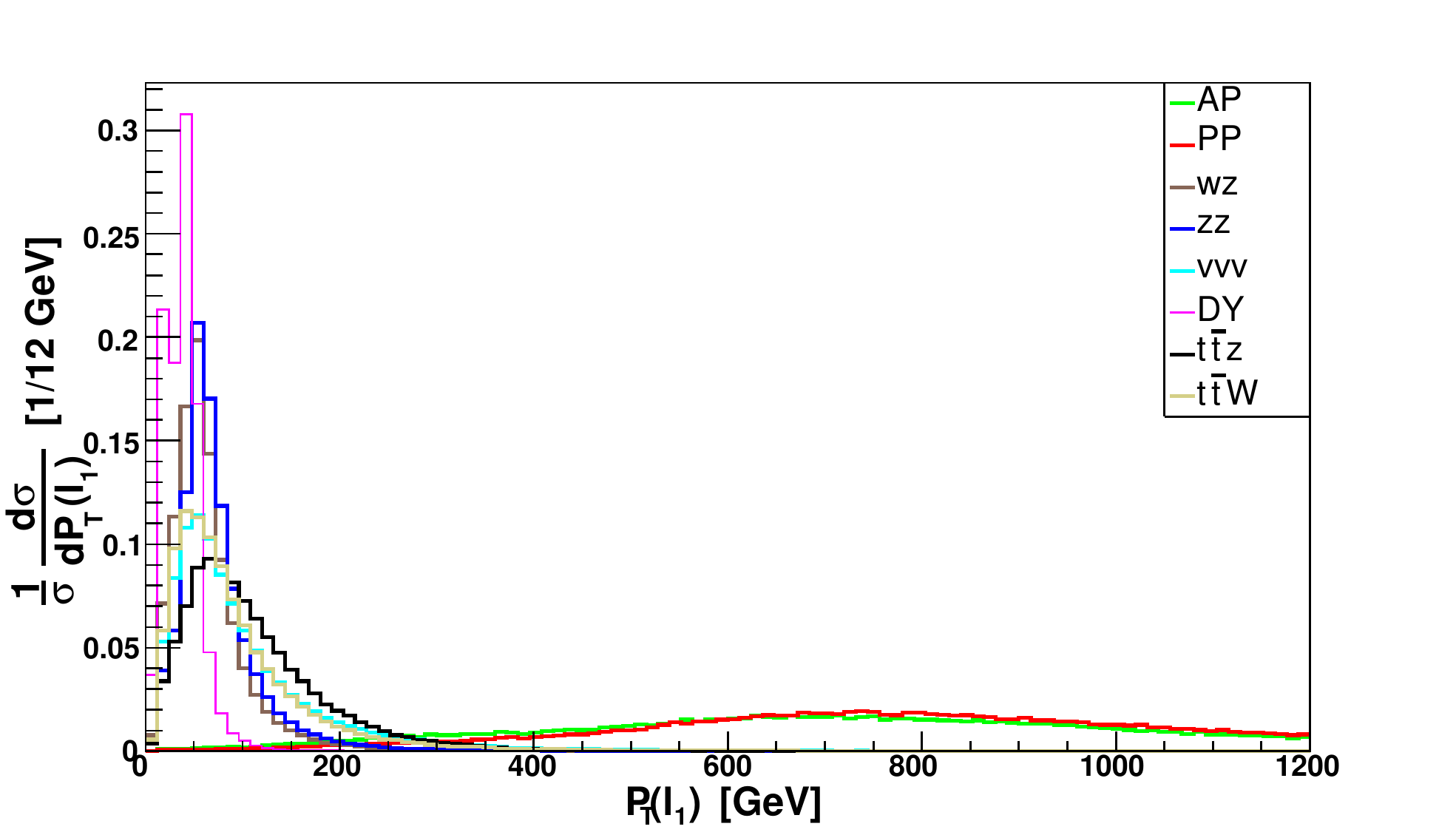}%
 	\includegraphics[width=9cm,height=5.5cm]{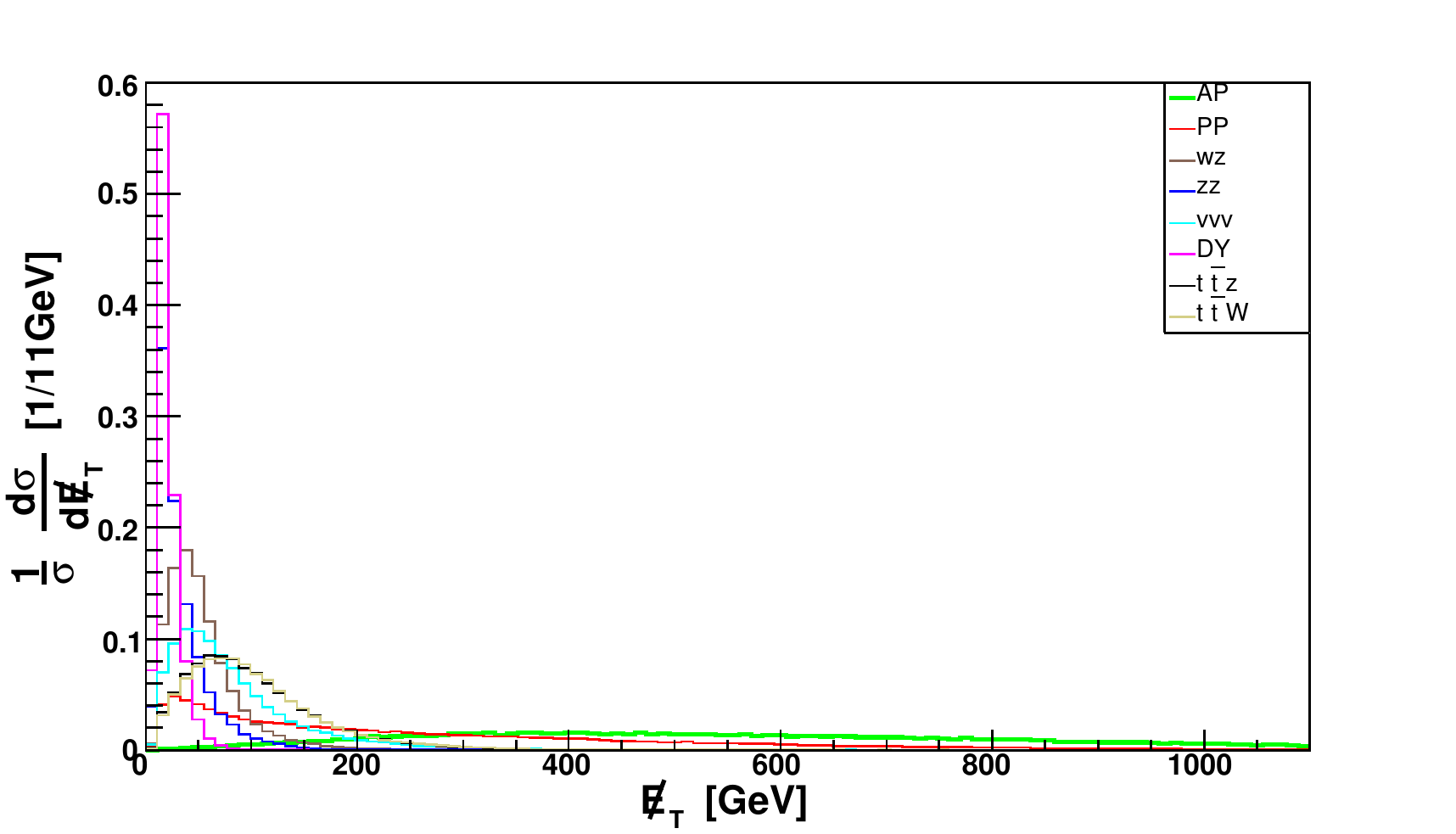}
 	\includegraphics[width=9cm,height=5.5cm]{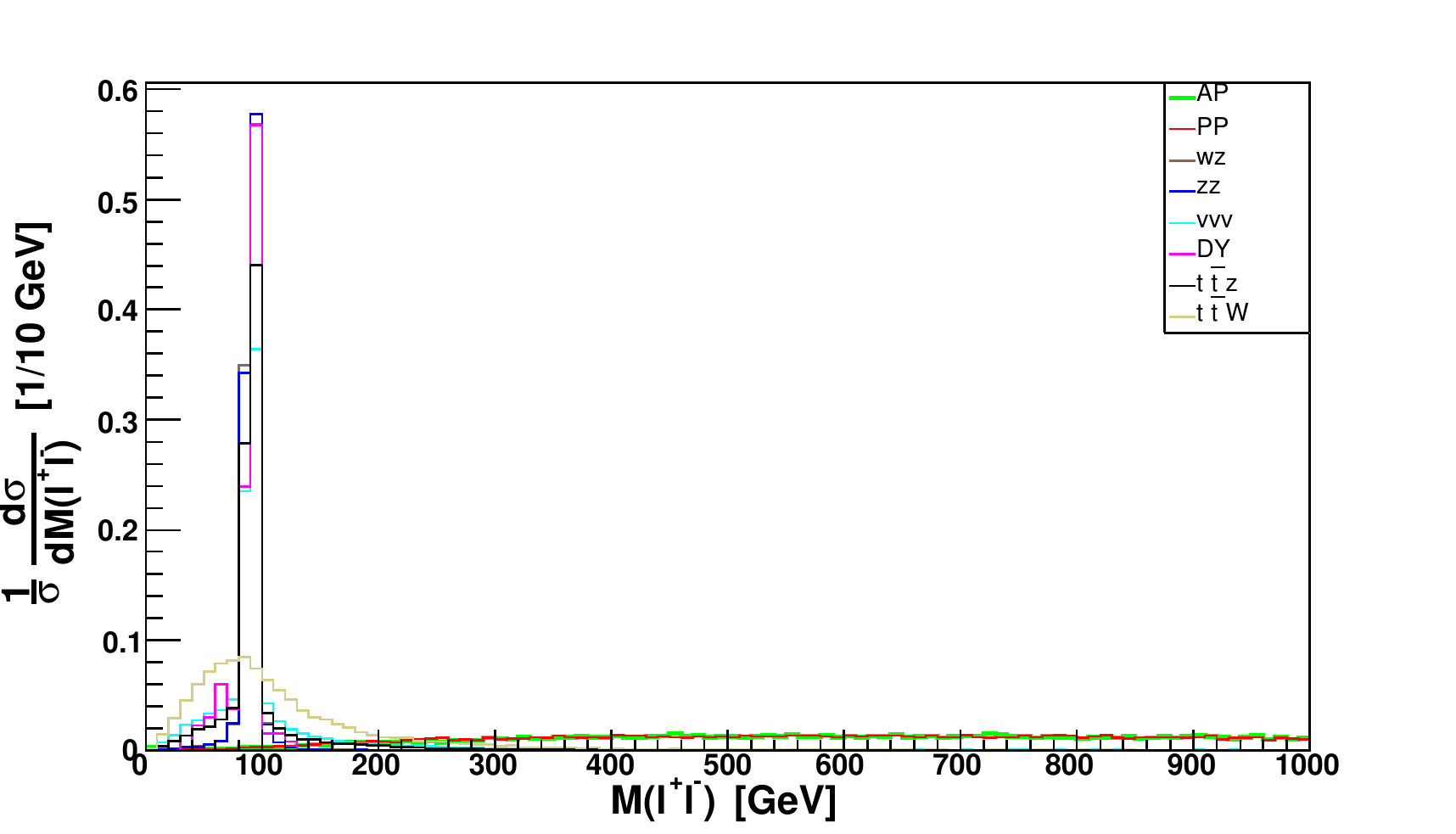}%
 	\includegraphics[width=9cm,height=5.5cm]{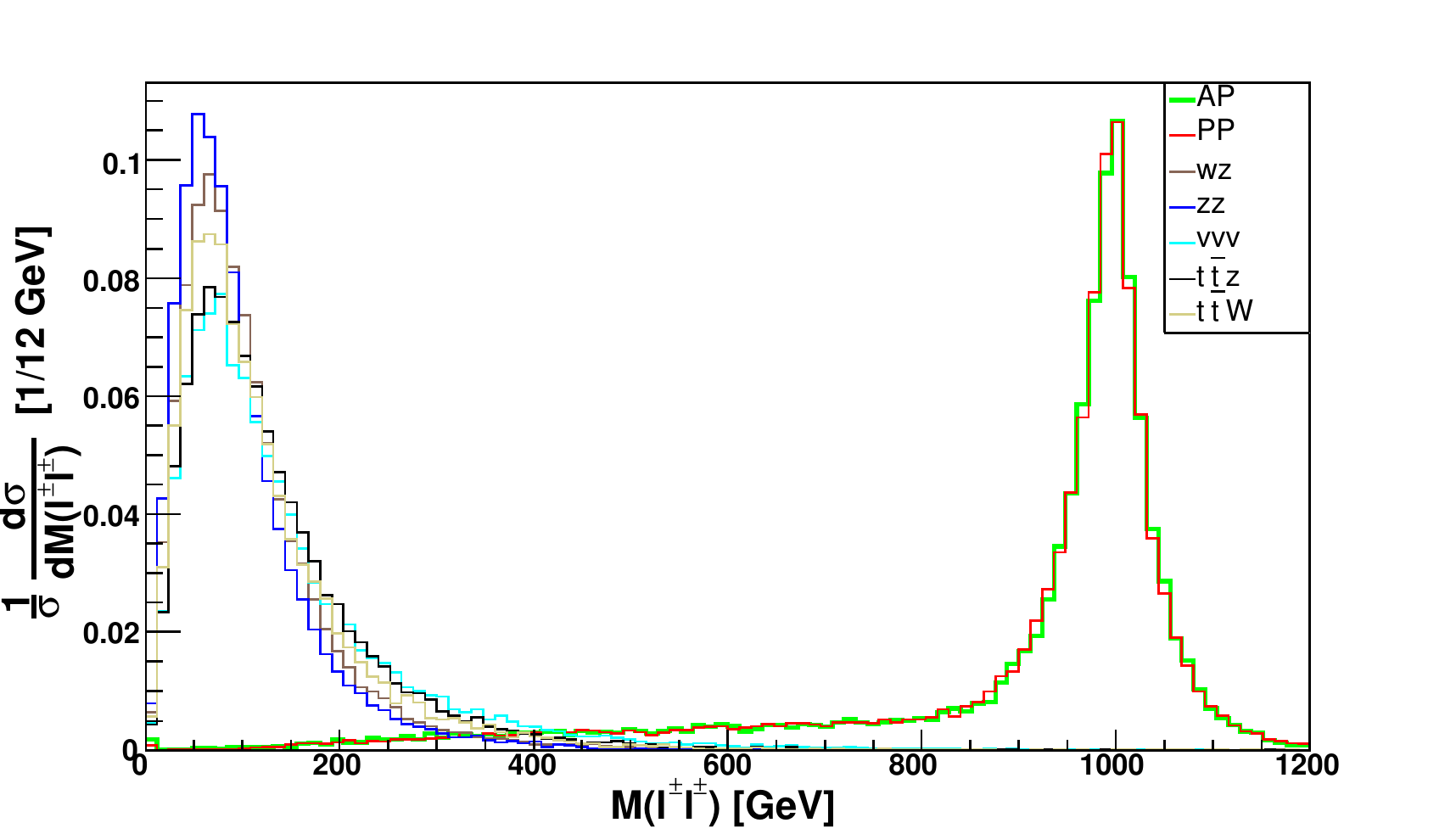}
 	
 	\caption{Normalised distribution of transverse momentum of leading lepton $P_T(l_1)$, missing transverse energy $\slashed{E}_{T}$, opposite sign di-lepton invariant mass $M(l^+l^-)$, same-sign di-lepton invariant mass $M(l^{\pm}l^{\pm})$ for both $3l$ signal and background events.}
 	\label{fig:dist3l}
 \end{figure}

To simulate the signal and backgrounds, we generate events in  MadGraph5, where we are applying the basic  cuts  $P_{T}(l) > 10$ GeV, $|\eta(l)| < 2.5$, $\Delta R_{ll} > 0.4$. Here we consider the same  neutrino mass pattern and  oscillation parameters that we have considered for $4l$ signal analysis in the  previous subsection. For this set of parameters, $H^{\pm}$ branching ratios are $e^{\pm}\nu\ : \ \mu^{\pm}\nu \ : \tau^{\pm}\nu = 0.48: 0.28: 0.24 $. {As the branching into muon and tau are very similar, we consider all the leptonic states into our analysis. }We consider the mass of charged Higgs as $1$ TeV and show distribution of transverse momentum of leading lepton $P_T(l_1)$, missing transverse energy $\slashed{E}_{T}$, opposite sign di-lepton invariant mass $M(l^+l^-)$, same-sign di-lepton invariant mass $M(l^{\pm}l^{\pm})$ in Fig.~\ref{fig:dist3l}. The two contributions to the signal events i.e.,  AP and PP are plotted separately. Most of the backgrounds events are distributed in the region of $P_T(l_1) \leq 200$ GeV, as   the leptons in backgrounds originate  from SM particles, but not from a heavy resonance. The signal sample on the contrary shows a peak at $P_T(l_1)> 200$ GeV. {A comparison in  the $\slashed{E}_{T}$ distribution between signal and background shows,  most of the background events  contain $\slashed{E}_{T}<200$ GeV.  Signal events coming from AP channel contain more $\slashed{E}_{T}$, as  compared to that of PP channel. This occurs, as in  AP channel, $\slashed{E}_{T}$ results from direct decay of $H^{\pm}$ into a lepton and neutrino. For  PP channel, the source of $\slashed{E}_T$   is  the decay of $\tau$ into semi-leptonic/leptonic  final states, or the mis-measurement of jet energy.} {We also show other distributions, such as, opposite sign and same-sign di-lepton invariant mass distributions. } In the opposite sign di-lepton invariant mass distribution,  background events peak around $Z$ mass. In the same-sign di-lepton invariant mass distribution, signal events peak at a higher value of $M(l^\pm l^\pm)$ as they directly originate from $H^{\pm \pm}$. Most of the background events are accumulated in lower  $M(l^{\pm}l^{\pm})$ region.

 To suppress the backgrounds, we consider the following selection criteria:
 
 \begin{itemize}
  	
  	\item $A_1 \ :$ Number of leptons  $\  N_{l} = 3$. We demand exactly 3 isolated leptons in the final state. 
  	\item $A_2 \ : $ Sum of lepton charge is $\pm 1$,  \ $\abs{\Sigma l_{charge}}=1$. Charge configuration of leptons are either $++-$ or $--+$.
  \item $A_3 \ : $ Invariant mass of opposite sign leptons falls within a 10 GeV mass window around $M_Z$, \ $ \abs{M(l^+ l^-) - M_{Z}} > 10$ GeV. {This cut  effectively  removes most of the backgrounds, from   $Z$ decay.}
  	
  	\item $A_4 \ :$ The transverse momentum of leading lepton, $  \  P_{T}(l_1)\ge150 $ GeV. We implement  a 150 GeV cut on $P_T$ of leading lepton, as SM background events contain  soft lepton compare to that of signal events. 
  	\item $A_5 \ :$ The missing transverse energy, $\ \slashed{E}_{T} > 100 $ GeV. We collect  events with $\slashed{E}_{T} > 120 $ GeV.
  
 	\item $A_6 \ : 	$ Same-sign di-lepton invariant mass  $ \ \abs{M(l^\pm l^\pm) - M_{H^{\pm\pm}}} \leqslant 50$, i.e., events within 100 GeV are selected.
  	
  \end{itemize}
 
 In Table~\ref{fig:3ltable},  we show  signal and background cross-sections after applying each of the selection cuts.
 \begin{itemize}
  	
  	\item $c_1 \ : \  N_{l} = 3$.
  	\item $c_2 \ : c_1 \ \text{and} \ \abs{\Sigma l_{charge}}=1.$
  	
  	\item $c_3 \ : \ c_2 \ \text{and} \ \abs{M(l^+ l^-) - M_{Z}} > 10$ GeV.
  	\item $c_4 \ : \ c_3   \ \text{and} \ P_{T}(l_1)\ge150$ GeV. 
  	\item$c_5 \ : \ c_4   \ \text{and} \ \slashed{E}_{T} \ge 120 $ GeV.
	\item $c_6 \ : \ c_5   \ \text{and} \ \abs{M(l^\pm l^\pm) - M_{H^{\pm\pm}}} \leqslant 50$ GeV.
  	
  \end{itemize}
  
 The  partonic  cross-section for AP channel  is $0.4501$ fb,  before applying any cut. For PP channel, this  is $0.2683$ fb. Most of the backgrounds  have very large cross-sections as compared to the signal.  At the detector level,  
 demanding three leptons reduces  the cross-sections significantly. The $Z$-veto (cut $c_3$) and demanding  high $P_T$ of leading lepton (cut $c_4$)    removes many backgrounds.  The missing transverse energy cut ($c_5$) also helps to suppress  backgrounds. SM processes like Drell-Yan ($DY$), $ t \bar{t}$ which give $3l$  due to jet faking as lepton, are left with negligible cross-sections after applying above mentioned cuts. Therefore, we do not show them explicitly in the Table.~\ref{fig:3ltable}. Similarly the virtual photon contribution to $3l$ signal does not survive at the end. Cut $c_5$ reduces the AP and PP cross-section to one sixth and one ninth of its initial value.
Although, at the partonic level, signal cross-sections are very small compared to that of SM backgrounds, we suitably choose  selection criterion, that suppress most of the backgrounds and keep a significant number of signal events. SM background cross-section finally reduced to around $0.0015$ fb. For an integrated luminosity of $1000 \ \rm{fb}^{-1}$,  we get around $79$ and $30$ number of signal events for AP and PP channels, respectively.
Note that,  most of the backgrounds in both tri-lepton and four-lepton channel drop off, after including the invariant mass cuts. For completeness, we however consider all the backgrounds,  and show the effect of selection cuts.

\begin{table}[h]
	\begin{tabular}{|c|c|c|c|c|c|c|c|}
 			\hline
& \multicolumn{2}{c|}{	$\sigma$ (fb) for signal } &\multicolumn{5}{c|}{$\sigma$ (fb) for backgrounds} \\ \hline
&  AP  & PP &$\ WZ \ $   & $\ ZZ \ $ & $VVV$  &$ t \bar{t}W$&$t \bar{t}Z$  \\ \hline
before cut & 0.4501 & 0.2683& 702.333&83.5597 & 9.49& 20.38&14.4134\\\hline
after $c_1$ &0.1456& 0.1052& 195.368&22.104&2.48&3.868&4.637 \\ \hline
after $c_2$ & 0.1453&0.1051&195.2 &22.09&2.476&3.853&4.629\\ \hline
after $c_3$&0.14497&0.1048& 17.158&1.8943&1.174& 3.391&1.577 \\ \hline
after $c_4$ &0.14493&0.1047&0.899 & 0.122& 0.265&0.725&0.354    \\ \hline
after $c_5$ &0.14074& 0.0672&0.3582&0.0192&0.1138&0.3053&0.1503\\\hline		
after $c_6$ &0.0793&0.0308&$\approx$0&$\approx0$&0.0005&0.001&$\approx$0\\\hline		
 		\end{tabular}
	\caption{Backgrounds and signal cross-sections for $\sqrt{s}= 27$ TeV after the final selection cuts for $3l$ final state. For signal $M_{H^{\pm\pm}}=1$ TeV.}
	\label{fig:3ltable}
\end{table}

	\begin{figure}[b]
	\includegraphics[width=7.9cm,height=5cm]{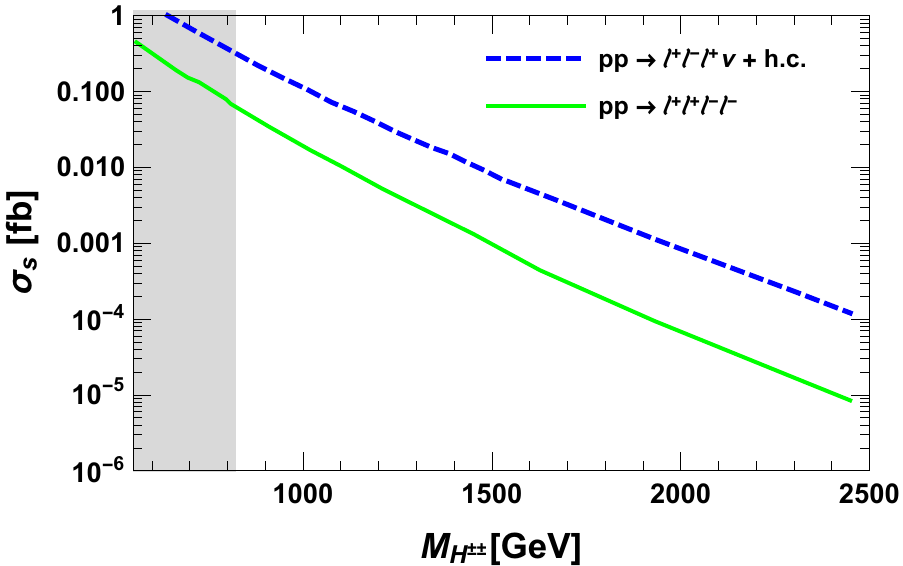}%
	\includegraphics[width=7.9cm,height=5cm]{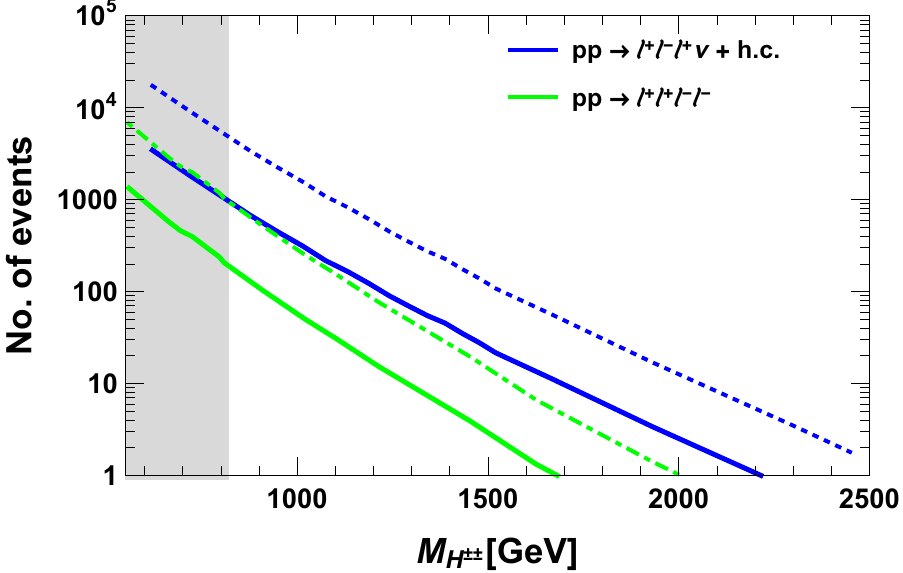}
	\includegraphics[width=7.9cm,height=5cm]{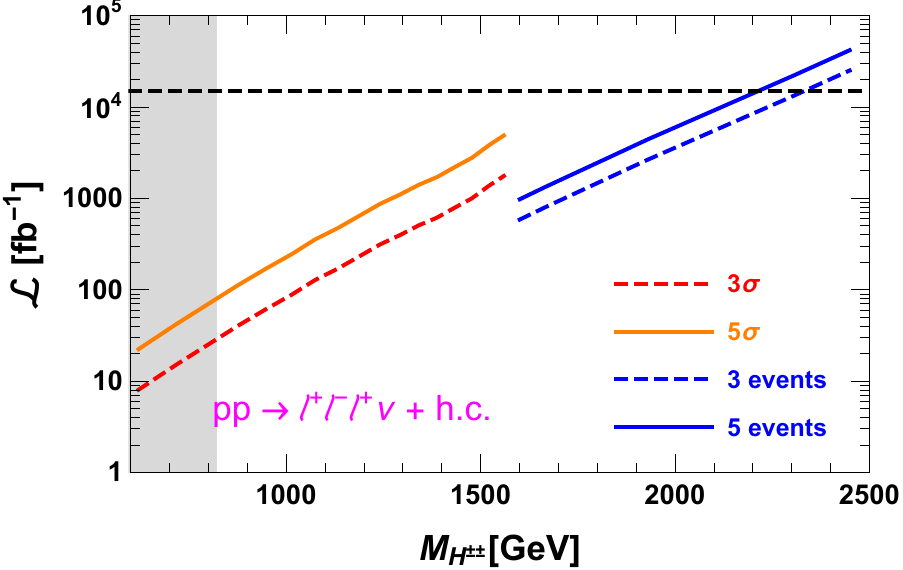}%
	\includegraphics[width=7.9cm,height=5cm]{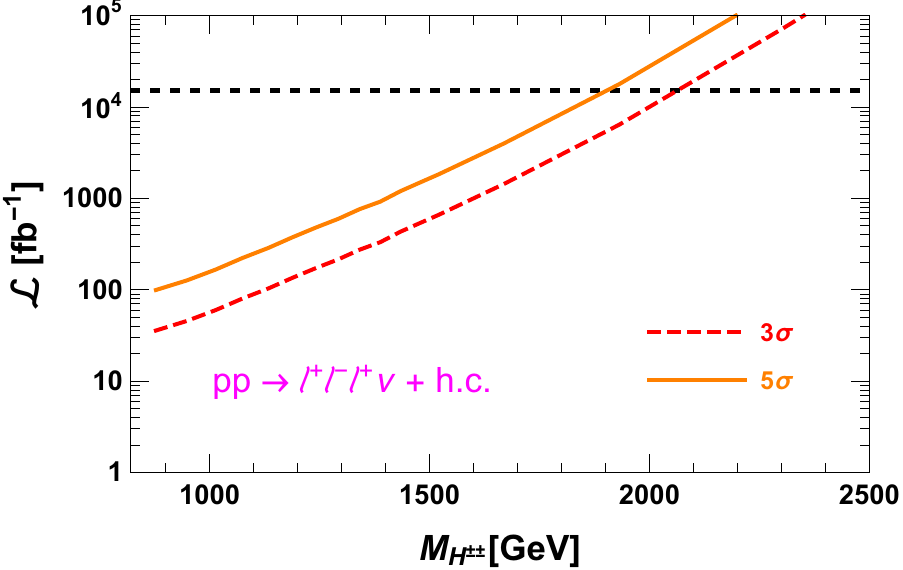}
	\includegraphics[width=7.9cm,height=5cm]{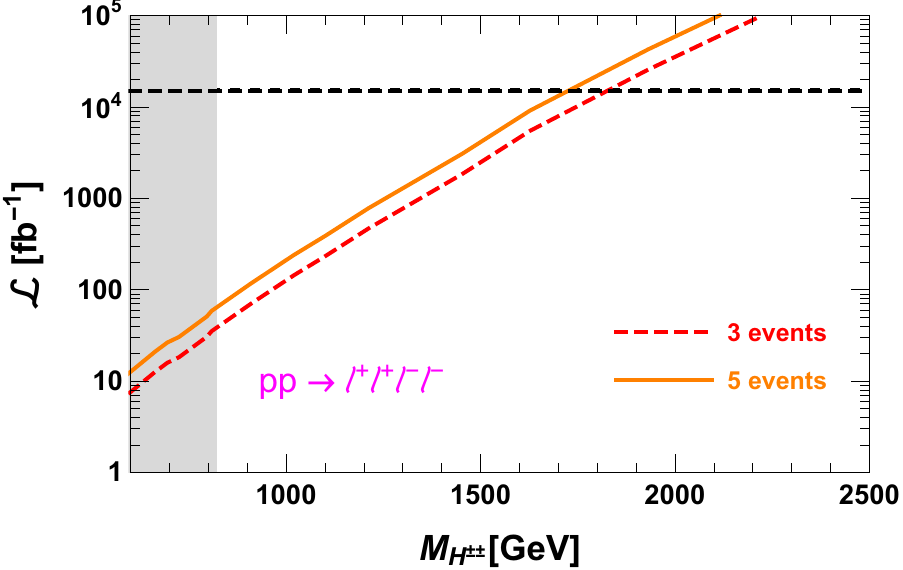}	
	\caption{ Upper panel: variation of tri and four lepton cross-sections after all cuts (left plot)  and number of events (right plot) for 3 $\text{ab}^{-1}$ (solid line), and 15$\, \text{ab}^{-1}$ (dashed, dot-dashed  line) luminosity as a function of $M_{H^{\pm\pm}}$. Middle panel:  variation of  required luminosity to reach $3\sigma$ and $5\sigma$ significance, and number of events $N=3,5$  versus $M_{H^{\pm\pm}}$ for tri-lepton signal. Lower panel: variation of  required luminosity to observe  number of events $N=3,5$ versus $M_{H^{\pm\pm}}$ for four-lepton signal. The gray shaded band represents the excluded region from CMS search~\cite{CMS-PAS-HIG-16-036}.}
	\label{fig:27_pred}
\end{figure}
 
We show  the variation of effective cross-sections of the  $4l$ and $3l$ final states  with $M_{H^{\pm\pm}}$  in Fig.~\ref{fig:27_pred}. The green line  in  top left plot represents the cross-section for $4l$ final state. The blue-dashed line denotes the cross-section for $3l$ final state, taking into account both pair and associated production. Although the dominant contribution for $3l$ final state comes from associated production of $H^{\pm\pm}$, pair production channel also contributes to this by a significant amount. In top right plot,  we show the variation of number of events as a function of $M_{H^{\pm\pm}}$. We consider two different luminosities 3 $\text{ab}^{-1}$, and the projected 15 $\text{ab}^{-1}$ of  HE-LHC.

 Finally, we calculate the statistical significance for the tri-lepton  channel:  \begin{equation} \mathcal{S} =  \dfrac{s}{\sqrt{s+b}}. \end{equation}
In the above,  $s$  and  $b$ are the number of signal and background events  after all of the above mentioned selection cuts. The required luminosity ($\mathcal{L}$) to achieve a desired significance ($\mathcal{S}$) therefore scales as 
 \begin{equation}
\sqrt{ \mathcal{L}}=\mathcal{S} \dfrac{\sqrt{\sigma_b+\sigma_s}}{\sigma_s}.
 \label{lum1}
 \end{equation}
 where $\sigma_s$ and $\sigma_b$ are the signal and background cross-sections after all the cuts. We obtain,  $3.3\sigma$ significance for a 1 TeV doubly charged Higgs with 100 $\text{fb}^{-1}$ luminosity, in tri-lepton channel. After including systematic uncertainty\footnote{We consider 2-6$\%$ uncertainty for the lepton identification~\cite{CMS-PAS-HIG-16-036} and $5\%$ uncertainty  on missing transverse enery scale~\cite{Sirunyan:2018zut}.} the significance decreases to $3.2\sigma$. For the four-lepton final states, and for tri-lepton channel in  higher mass range, there is no  SM background. This happens due to the very high invariant mass cut of the same-sign di-lepton. Therefore, 
 for these cases, we simply scale the required luminosity as 
 \begin{equation}
 \mathcal{L}=\dfrac{N}{\sigma_s}, 
 \label{lum2}
 \end{equation}
  where $N$ is the number of signal events, and $\sigma_s$ is the cross-section after all cuts. The two plots in the middle panel show the required luminosity to observe  tri-lepton states for the $M_{H^{\pm \pm}}$ in between 820-2500 GeV. In the left plot, we impose   a flat 100 GeV window of  invariant mass  $l^{\pm} l^{\pm}$ around $M_{H^{\pm \pm}}$. The orange and red lines have been obtained by using Eq.~\ref{lum1}, where the background is
 sizeable. In the higher mass range, there is almost no background. We therefore use Eq.~\ref{lum2}. $H^{\pm\pm}$ with mass $\sim 2.2$ TeV can be discovered with 5 number of events. We also present a conservative estimate in the right plot (middle panel), where we impose a cut on same-sign di-lepton invariant mass $M(l^{\pm} l^{\pm} ) > 820$ GeV, for which we have a constant background 0.008 fb. This shows that $\sim$ 2 TeV $M_{H^{\pm \pm}}$ can be discovered in the tri-lepton channel with integrated luminosity 15 $\rm{ab}^{-1}$.  For the four-lepton final state (lower panel), we find that approximately five events can be observed for $M_{H^{\pm \pm}}=1.7$ TeV  with the same luminosity.\\

 The analysis that we present here, depends on the chosen final state, branching ratios, as well as, the selection cuts. Additionally, the results presented are at the LO level. However, we would like to present an estimate of the results taking into account K-factor for leading background. For signal K-factors at 14 TeV and 100 TeV LHC have been calculated in Ref.~\cite{Fuks:2019clu},  which are not widely different. For background cross-sections  we calculate the NLO corrections at 27 Tev LHC using MadGraph5\_aMC@NLO(v2.6)~\cite{Alwall:2014hca}. K-factor for $t\bar{t}W$ ($ VVV$) is 1.86 (2.2). Considering the LO cross-sections,  the statistical significance ($\mathcal{S}$) for $3l$ signal ( with $M_{H^{\pm\pm}}=1$ TeV and $\mathcal{L}=100 \  \text{fb}^{-1}$) is 3.29. If we include K-factors the significance will increase to 3.6. Including NLO corrections to the pair-production of doubly charged Higgs ($K=1.25$), the discovery reach of ${H^{\pm \pm}}$ in the four-lepton final state   can extend upto  $1.8$ TeV. We also cross-checked the discovery potential of a very heavy $H^{\pm\pm}$ at HE-LHC.  We would like to point out  that for a very large mass, such as, $M_{H^{\pm \pm}}=4$ TeV, the pair-production  cross-section via DY process drops to  $\sigma  \sim 3 \times 10^{-5}$ fb.  Folded with  approximately 50$\%$ branching ratios (note that, branching ratio in any of the leptonic final states can not be $100\%$), the final cross-section for four-lepton  channel becomes  very small. Due to this, we do  not obtain any signal event with 15 $\text{ab}^{-1}$ luminosity for 4 TeV doubly charged Higgs mass.

 \section{Conclusion \label{conclu}}
 
 We analyse discovery prospect of a doubly charged Higgs -  a particle content of type-II seesaw model in $p p $ collider (HE-LHC). We focus on the region of small triplet vev, where the doubly and singly charged Higgs naturally decays to 
 same-sign di-leptons and  a charged lepton and neutrino final states, respectively. We analyse in detail multi-lepton signatures, containing tri and tetra-leptons in the final states. The model signatures in this low vev regime strongly depends on the neutrino oscillation parameters, neutrino mass hierarchy, and the lightest neutrino mass scale.  We perform a robust estimation of the maximal possible branching ratios, that each of the leptonic modes can accommodate. 
 The constraint on doubly charged Higgs mass from  individual  leptonic channel somewhat weakens, after taking into account correct branching ratios.   The doubly charged Higgs and singly charged Higgs couple to the leptons through the same Yukawa coupling. We explore the relation between the   branching ratios of  singly and doubly charged Higgs decays.  Our major findings are,

  \begin{itemize}
  
  \item
  The branching of doubly charged Higgs  into same-sign leptons is augmented with large variation due to the uncertainty  of neutrino oscillation parameters. We find that in IH neutrino mass spectrum, and among all the leptonic decay modes of $H^{\pm \pm}$,  $e^{\pm}e^{\pm}$ mode is least uncertain for the entire range of lightest neutrino mass $m_0$. This decay mode predicts a  lower value of branching ratio, which is $\text{BR}(H^{\pm \pm} \to e^{\pm} e^{\pm})> 0.015$. Therefore, observation of doubly charged Higgs in any other leptonic decay mode except $ e^{\pm} e^{\pm}$ with an upper limit on branching ratio $\text{BR}(H^{\pm \pm} \to e^{\pm} e^{\pm})< 0.015$ will rule out IH. The singly charged Higgs decay to $H^{\pm} \to e^{\pm} \nu$ is the least uncertain among all charged Higgs decays, with a predicted branching ratio, that vary between  $ \sim33\%- 50\%$ for the variation of the lightest neutrino mass $m_0 \sim 10^{-4}\, {\rm{eV}} -1$ eV. 
  \item The interaction of $H^{\pm\pm}$ and $H^{\pm}$ with leptons are governed by the same Yukawa couplings, and hence their leptonic branching ratios are ralated. In IH, for a fixed $m_0$, and for a fixed value of $\text{BR}(H^{\pm \pm} \to e^{\pm} e^{\pm})$, there is a small variation in  $\text{BR}(H^{\pm} \to e^{\pm} \nu)$.  Similar result also exists between $\text{BR}(H^{\pm \pm} \to e^{\pm} \mu^{\pm})$ and $\text{BR}(H^{\pm} \to e^{\pm} \nu)$.
\item 
We perform a detailed  analysis to find out discovery prospect of  tri and tetra-lepton final states at a future $p p $ collider, that can operate with c.m.energy $\sqrt{s}=27 \, \rm{TeV}$.  We consider both the pair and associated productions, and a benchmark point of $H^{\pm \pm}$ with a mass  1 TeV. In tetra-lepton final state, we find using 1000 $\rm{fb}^{-1}$ integrated luminosity, 19 events can be observed.  In tri-lepton final state, we find that the same mass can be discovered at 27 TeV LHC with a significance of  more than $5\sigma$  for 1000 $\rm{fb}^{-1}$ luminosity. Higher mass region of ${H^{\pm \pm }} $ can be probed  with  more luminosity. We find that sensitivity reach for ${H^{\pm \pm }}$ in tri-lepton channel is more compared to that in four-lepton channel, as both the pair and associated production of ${H^{\pm \pm }}$ contribute to the former. ${H^{\pm \pm }} $ upto mass $\sim 2.2$ TeV can be probed in tri-lepton channel with  $15 \ \text{ab}^{-1}$ integrated luminosity. In four-lepton channel, five events can be observed for $M_{H^{\pm \pm}}$ $\lesssim$ $1.7$ TeV with the same luminosity.   
\end{itemize}

 \section{Acknowledgement}
 
 MM acknowledges the DST-INSPIRE Grant (IFA-14-PH-99). RP acknowledges the support of IOP Cluster SAMKHYA. 
	\section{Appendix\label{app}}
	
	Here we expand  neutrino mass matrix in terms of the PMNS mixing angles, CP phases, and the mass square differences \cite{Perez:2008ha}.  We first consider the $H^{\pm \pm}$ decay and then $H^{\pm}$ decay. 

	\begin{itemize}
	
		\item  
		\underline{$H^{\pm\pm} \ \text{Decays}$}\\ 
		
		The branching ratio has the following form 
		\begin{equation}
{\rm{BR}}(H^{\pm \pm}\to l_i^\pm l_j^\pm) =  \frac{\Gamma_{l_il_j}}{\sum_{kl}	\Gamma_{l_kl_l}} = \frac{2}{(1+\delta_{ij})} \frac{\abs{Y^\nu_{ij}}^2}{\sum_{kl}\abs{Y^\nu_{kl}}^2 }, \end{equation} 
where 
\begin{equation}
\quad \sum_{kl}\abs{Y^\nu_{kl}}^2 = \dfrac{1}{2v^2_\Delta}\sum_i m_i^2.
\end{equation} 
The Yukawa and neutrino mass matrix are related as,
		\begin{equation}
			Y^\nu = \dfrac{M^\nu}{\sqrt{2}v_\Delta}=\dfrac{V^*_{\text{PMNS}} m^\nu_{\text{d}}V^\dag_{\text{PMNS}}}{\sqrt{2}v_\Delta}.
		\end{equation}
	
	Neutrino mass matrix elements can be written as a function of light neutrino masses and mixing parameters in the following forms (Here $c_{ij}\equiv \text{cos}\theta_{ij} \ \text{and} \ s_{ij}\equiv \text{sin}\theta_{ij}$):
	\begin{eqnarray}
	\non
	M^\nu_{11} &=&
	m_1 e^{-i \Phi_1} c_{12}^2 c_{13}^2
	+m_2 s_{12}^2 c_{13}^2
	+m_3 e^{i(2 \delta -\Phi_2)} s_{13}^2
	\\ \non
	M^\nu_{22} &=&
	m_1 e^{-i \Phi_1} \left(-s_{12} c_{23} -e^{-i \delta } c_{12} s_{13} s_{23}\right){}^2
	+m_2 \left(c_{12} c_{23}-e^{-i \delta } s_{12} s_{13} s_{23}\right){}^2
	+m_3e^{-i \Phi_2} c_{13}^2 s_{23}^2
	\\ \non
	M^\nu_{33} &=&
	m_1 e^{-i \Phi_1} \left(s_{12} s_{23}-e^{-i\delta } c_{12} s_{13} c_{23}\right){}^2
	+m_2 \left(-c_{12} s_{23}-e^{-i \delta } s_{12} s_{13} c_{23}\right){}^2
	+m_3 e^{-i \Phi_2} c_{13}^2 c_{23}^2
	\\ \non
	M^\nu_{12} &=&
	m_1 e^{-i \Phi_1} c_{12} c_{13} \left(-s_{12} c_{23}-e^{-i \delta } c_{12} s_{13} s_{23}\right)
	+m_2 s_{12} c_{13} \left(c_{12} c_{23}-e^{-i \delta } s_{12} s_{13} s_{23}\right)
	\\ \non
	&+&m_3 e^{i (\delta -\Phi_2)} s_{13} c_{13} s_{23}
	\\ \non
	M^\nu_{13} &=&
	m_1 e^{-i \Phi_1} c_{12} c_{13} \left(s_{12} s_{23}-e^{-i \delta } c_{12} c_{23} s_{13}\right)
	+m_2 c_{13} s_{12} \left(-c_{12} s_{23}-e^{-i\delta } s_{12} s_{13} c_{23}\right)
	\\ \non
	&+&m_3 e^{i (\delta -\Phi_2)} s_{13} c_{13} c_{23}
	\\ \non
	M^\nu_{23} &=&
	m_1 e^{-i \Phi_1} \left(s_{12} s_{23}-e^{-i \delta } c_{12} s_{13} c_{23}\right)
	\left(-s_{12} c_{23}-e^{-i \delta } c_{12} s_{13} s_{23}\right)
	\\ \non
	&+&m_2 \left(-c_{12} s_{23}-e^{-i \delta } s_{12} s_{13} c_{23}\right)
	\left(c_{12} c_{23}-e^{-i \delta } s_{12} s_{13} s_{23}\right)
	+m_3 e^{-i \Phi_2} c_{13}^2 s_{23} c_{23}
	\end{eqnarray}
	
	Following  \cite{deSalas:2017kay}, we consider the following set of  3$\sigma$ variation of  PMNS  mixing angles and mass squared differences:
	
	$ \sin^{2}\theta_{12} \Rightarrow [0.273-0.379]
	\ $ \\
	$ \sin^2 \theta_{23}   \Rightarrow   [0.445 - 0.599]_{\text{NH}} ,\quad [0.453-0.598]_{\text{IH}}$ \\ 
	$\sin^2\theta_{13}  \Rightarrow ( [1.96 - 2.41]  \times 10^{-2})_{\text{NH}} , \quad ([1.99-2.44] \times 10^{-2})_{\text{IH}}$	

	$\Delta m_{21}^2   \Rightarrow [7.05-8.14] \ \times 10^{-5} \  \text{eV}^2$\\
$	|\Delta m_{31}^2 |  \Rightarrow  \   [2.41-2.60]  \ \times \  10 ^{-3}\  \text{eV}^2  \ \  \text{(NH)}$\\
$|\Delta m_{31}^2 |  \Rightarrow  \    \  [2.31-2.51] \ \times10^{-3} \ \text
		{eV}^2  \  \text{(IH)}$\\
		
		Below, we analytically calculate the maximum value of branching ratios for different leptonic decay modes of $H^{\pm\pm}$. For this calculation we assume those values of the oscillation parameters that gives a maximum branching ratio for a given decay mode. The set of parameters is not necessarily same for each mode.	
		
\begin{enumerate}
\item{ Inverted Hierarchy neutrino mass spectrum ($m_3\approx0$):} Here, due to very small mass splitting between $m_{1}$ and $m_{2}$, 
$m_1\approx m_2=m$. This gives summation of light neutrino mass as  $ \sum_{j} m_j^2=2 m^2$. We identify different phases, for which the  branching ratios in $ee$, $\mu \mu, e \mu $ modes are maximal. 

\begin{itemize}
	\item
	
For  $\phi _1=0$, the $(1,1)$ element of neutrino mass matrix $ M^\nu_{11}= m c_{13}^2$. The branching ratio is $ \text{BR}\left(H^{\pm\pm }\to  e ^{\pm } e ^{\pm }\right)
 \approx \dfrac{c_{13}^4}{2}\approx 0.44$. 
 
\item For  $\delta =\phi _1=\phi _2=0$, the $(2,2)$ element of neutrino mass matrix is  $ M^\nu_{22}=\left(c_{23}^2 + s_{13}^2 s_{23}^2\right)m$. This gives the maximal  branching ratio as
$\text{BR}\left(H^{\pm  \pm }\to  \mu ^{ \pm } \mu ^{\pm }\right)\approx \dfrac{c_{23}^4}{2} \approx 0.18$. 

\item  For $ \phi _2=\pi ,\delta =0$, the $(1,2)$ element of neutrino mass matrix is   $M^\nu_{12}=c_{12} c_{13} m_1 \left(c_{23} s_{12}+c_{12} s_{13} s_{23}\right)+c_{13} m_2 s_{12} \left(c_{12} c_{23}-s_{12} s_{13} s_{23}\right)$. The branching ratio in $e\mu$ mode is   $$\text{BR}\left(H^{\pm  \pm }\to e^{ \pm } \mu ^{\pm }\right)\approx 2c^2_{13} \dfrac{\left((c_{12}^2-s_{12}^2) s_{13} s_{23}+2 c_{23} c_{12} s_{12}\right)^2}{2} \approx 0.48.$$
		
		\end{itemize}

			\item{ Normal Hierarchy neutrino mass spectrum ($m_1\approx 0$):}  The two other masses are 
$m_2\approx0.2 \ m_3$. 
		\begin{itemize}
		\item	 For  $2\delta-\phi_2=0$, the $(1,1)$ element of neutrino mass matrix is  $ M^\nu_{11}= m _2s_{12}^2c_{13}^2+ m_3 s_{13}^2$, that gives $ \text{BR}\left(H^{\pm\pm }\to  e ^{\pm } e ^{\pm }\right)\approx 0.008 $. 
		\item For  $\delta =\pi$ and $\phi _2=0$, the $(2,2)$ element of neutrino mass matrix is $ M^\nu_{22} = c_{12}^2 c_{23}^2 m_2+c_{13}^2 m_3 s_{23}^2 + 2 c_{12} c_{23} m_2 s_{12} s_{13} s_{23}+m_2 s_{12}^2 s_{13}^2 s_{23}^2$. Ignoring the  3rd and 4th terms, the branching ratio in $\mu\mu$ becomes  $$ \text{BR}\left(H^{\pm  \pm }\to  \mu ^{ \pm } \mu ^{\pm }\right)\approx \dfrac{(c_{12}^2 c_{23}^2 m_2+c_{13}^2 m_3 s_{23}^2)^2}{m^2_3} \approx 0.4.$$
		\item
			For $\phi _2=\delta=\pi$, the (1,2) element of neutrino mass matrix is  $ M^\nu_{12}=c_{13} m_3 s_{13} s_{23}+c_{13} m_2 s_{12} \left(c_{12} c_{23}+s_{12} s_{13} s_{23}\right)$. The branching ratio is 
			$$\text{BR}\left(H^{\pm  \pm }\to e^{ \pm } \mu ^{\pm }\right)\approx \dfrac{2\left(0.2 c_{12} c_{23} m_3 s_{12}+s_{13} s_{23} \left(0.2 m_3 s_{12}^2+m_3\right)\right)^2}{m^2_3}\approx 0.072.$$
		\end{itemize}
		The different ratios of branching ratio are \\ \\
			$\dfrac{\text{BR}^{\max}(H^{\pm \pm }\to e^{\pm } \mu^{\pm })_{\text{IH}}}{\text{BR}^{\max}(H^{\pm \pm }\to e^{\pm } \mu^{\pm })_{\text{NH}}}
			\approx\dfrac{\left(c_{12}^2 s_{13} s_{23}+2 c_{23} c_{12} s_{12}-s_{12}^2 s_{13} s_{23}\right)^2}{2 \left(0.2 c_{12} c_{23} s_{12}+s_{13} s_{23} \left(0.2  s_{12}^2+1\right)\right){}^2}\approx6.6$,\\
			
			$
			\dfrac{\text{BR}^{\max}(H^{\pm \pm }\to e^{\pm } e^{\pm })_{\text{IH}}}{\text{BR}^{\max}(H^{\pm \pm }\to e^{\pm } e^{\pm })_{\text{NH}}}
			\approx\dfrac{ c_{13}^4}{2\left(0.2 c_{13}^2 s_{12}^2+s_{13}^2\right)^2}\approx50$,\\
			
	$\dfrac{\text{BR}^{\max}(H^{\pm \pm }\to \mu^{\pm } \mu^{\pm })_{\text{IH}}}{\text{BR}^{\max}(H^{\pm \pm }\to \mu^{\pm } \mu^{\pm })_{\text{NH}}}
			\approx \dfrac{c_{23}^4}{2 (c_{12}^2 c_{23}^2 0.2+c_{13}^2 s_{23}^2)^2}\approx0.45$.\\
		\end{enumerate}
					Similarly, one can obtain such ratios for all other modes.

			\item \underline{$H^{\pm}$ Decays}\\
			
The coupling through which $H^{\pm}$ interact with charged lepton and neutrino  is 
			\begin{equation}
			Y^+ = \cos\theta^+ \frac{m^\nu_{d}V_{PMNS}^\dagger}{v_\Delta}.\end{equation}
			In the above,  $\theta^+$ is the singly charged Higgs mixing angle. The branching ratio has the following form, 
		\begin{equation}
			 {\text{BR}\left(H^{\pm }\to  l_j^{ \pm }\nu \right)\equiv \sum_{i=1}^{3} {\rm{BR}}(H^{\pm}\to l_j^\pm \nu_i) =\dfrac{X_j}{\sum_{i}^3 m_i^2}}, \quad   (l_j=e,\mu,\tau)\end{equation}
			 
where,  $ X_j $ is defined as, $X_j=\dfrac{v^2_\Delta}{\text{cos}^2\theta^+}\sum_{i=1}^{3}|Y^+_{ij}|^2. $  The $X_{1,2,3}$ has the following form, 
			\begin{eqnarray}
			\non 
			X_1 &=& m_1^2 c_{12}^2 c_{13}^2 + m_2^2 c_{13}^2  s_{12}^2+m_3^2 s_{13}^2. \\ \non
			X_2 &=& m_1^2 c_{23}^2 s_{12}^2+2 \text{cos}(\delta)  \left(m_1^2-m_2^2\right) c_{12} c_{23}s_{12} s_{13} s_{23}\\ \non &+&\left( m_3^2 c_{13}^2 +m_2^2 s_{12}^2 s_{13}^2\right) s_{23}^2+c_{12}^2 \left( m_2^2 c_{23}^2 +m_1^2 s_{13}^2 s_{23}^2\right).\\ \non
			X_3&=& m_3^2 c_{13}^2 c_{23}^2 -2 \text{cos}(\delta ) \left(m_1^2-m_2^2\right)c_{12} c_{23} s_{12} s_{13} s_{23}\\ \non 
			&+& s_{12}^2 ( m_2^2 c_{23}^2  s_{13}^2+m_1^2 s_{23}^2)+c_{12}^2 (m_1^2 c_{23}^2  s_{13}^2+m_2^2 s_{23}^2).
			\\ \non	
			\end{eqnarray}
Maximum value of branching ratio for $H^{\pm}\to l_j^{ \pm }\nu$ is presented bellow assuming different type of neutrino mass spectrum. 
\begin{enumerate}
\item Inverted Hierarchy neutrino mass spectrum ($m_3\approx0$):\\ Here, 
$m_1\approx m_2=m\Rightarrow \sum_{j} m_j^2=2 m^2$.
\begin{itemize}
\item
$X_1=c_{13}^2 m^2$ gives $\text{BR}\left(H^{\pm }\to  e^{ \pm }\nu \right)= \dfrac{c_{13}^2}{2}\approx 0.49$. Since there is only $\theta_{13}$ dependency in $X_1$, and $\theta_{13}$ is very well measured, therefore  $H^{\pm }\to  e^{ \pm }\nu$ decay modes has very less uncertainty. 
\item For $\delta=0$, $X_2$ has the following form $X_2=\left(1-s_{23}^2 \left(1+s_{13}^2\right)\right)m^2$, that  gives $ \text{BR}\left(H^{\pm }\to  \mu^{ \pm }\nu \right)= \dfrac{1-s_{23}^2 \left(1+s_{13}^2\right)}{2} \approx0.3.$
\item For $\delta=\pi$, $X_3$ has the following form $X_3=\left(s_{23}^2\left(1-s_{13}^2\right)+s_{13}^2\right)m^2$, that gives $ \text{BR}\left(H^{\pm }\to  \tau^{ \pm }\nu \right)= \dfrac{s_{23}^2\left(1-s_{13}^2\right)+s_{13}^2}{2}\approx 0.3.$\\
Note that, $H^{\pm }\to  \mu^{ \pm }\nu/ \tau^{\pm}\nu$ decay modes have nearly equal uncertainty as both depend on  $\theta_{23}$
and $\theta_{13}$ . 
\end{itemize}
\item Normal Hierarchy neutrino mass spectrum ($m_1\approx0$): For this spectrum $m_2=0.2 \ m_3$, that gives $ \sum_{j} m_j^2\approx m_3^2 $. The CP phase  $\delta =\pi (0)$, will maximize $X_2 (X_3)$ and hence branching ratios.
\begin{itemize}
\item$X_1\approx m_2^2 c_{13}^2  s_{12}^2+m_3^2 s_{13}^2$ gives  $\text{BR}\left(H^{\pm }\to  e^{ \pm }\nu \right)\approx c_{13}^2 (0.2)^2 s_{12}^2+ s_{13}^2 \approx 0.037$.
\item$X_2\approx m_3^2  c_{13}^2 s_{23}^2 $ gives $ \text{BR}\left(H^{\pm }\to  \mu^{ \pm }\nu \right)\approx c_{13}^2  s_{23}^2 \approx0.57$.
\item$X_3\approx  m_3^2 c_{13}^2 c_{23}^2$ gives $ \text{BR}\left(H^{\pm }\to  \tau^{ \pm }\nu \right)\approx c_{13}^2 c_{23}^2 \approx 0.53$.
	\end{itemize}

			\end{enumerate}

	\end{itemize}

 \bibliographystyle{utphys}
 \bibliography{t2seesaw.bib}

 \end{document}